%% file: PlenzBook-KozmaPuljicFreeman.tex
\documentclass{wileysev}
\usepackage{graphicx,subfigure}
\usepackage{dcolumn}
\usepackage{bm}
\usepackage{amsmath}
\usepackage{amssymb}
\usepackage{chapterbib}
\begin{document}

\offprintinfo{Criticality in Neural Systems}{Dietmar Plenz (ed.)}
\chapter{Thermodynamic Model of Criticality in the Cortex Based on EEG/ECoG Data}
\chapterauthors{Robert Kozma\affilmark{1}, Marko Puljic\affilmark{2, 1} and Walter J. Freeman\affilmark{3}
\chapteraffil{\affilmark{1}Center for Large-Scale Integrated Optimization and Networks (CLION) \\University of Memphis, Memphis, TN 38152, USA\\ \affilmark{2}Computational Science, Tulane University, New Orleans, LA 70118, USA\\ \affilmark{3}Division of Neurobiology, University of California at Berkeley, CA 94720, USA } 
}
\date{\today}

\prologue{... Thus logics and mathematics in the central nervous system, when viewed as languages, must structurally be essentially different from those languages to which our common experience refers. ... whatever the system is, it cannot fail to differ considerably from what we consciously and explicitly consider as mathematics.}{John Von Neumann \cite{vneumann58}\\
}

\vspace*{8mm}
\small{{\it Abstract} - Criticality in the cortex emerges from the seemingly random interaction of microscopic components and produces higher cognitive functions at mesoscopic and macroscopic scales. Random graphs and percolation theory provide natural means to describe critical regions in the behavior of the cortex and they are proposed here as novel mathematical tools helping us deciphering the language of the brain. }\\

\section{Introduction}

Von Neumann emphasized the need for new mathematical disciplines in order to understand and interpret the language of the brain \cite{vneumann58}. He has indicated the potential directions of such new mathematics through his work on self-reproducing cellular automata and finite mathematics. Due to his early death in 1957, he was unable to participate in the development of relevant new theories, including morphogenesis pioneered by Turing \cite{turing54} and summarized by Katchalsky \cite{katchalsky71}. Prigogine \cite{prigogine80} developed this approach by modeling the emergence of structure in open chemical systems operating far from thermodynamic equilibrium. He called the patterns dissipative structures, because they emerged by the interactions of particles feeding on energy with local reduction in entropy. Haken \cite{haken83} developed the field of synergetics, which is the study in lasers of pattern formation by particles, whose interactions create order parameters by which the particles govern themselves in circular causality. 

Principles of self-organization and metastability have been introduced to model cognition and brain dynamics \cite{hkb85}, \cite{jaskelso95}, \cite{haken02}. Recently the concept of self-organized criticality (SOC) has captured the attention of neuroscientists \cite{beggsplenz03}. There is ample of empirical evidence of cortex conforming to the self-stabilized, scale-free dynamics of the sand pile during the existence of quasi-stable states \cite{bak1996, beggs08, petermann09}. However, the model cannot produce the emergence of orderly patterns within a domain of criticality \cite{biolcyb05, freemanzhai09}. Bonachela \cite{bonachela10} describe SOC as Òpseudo-criticalÓ and suggest to complement self-organization with more elaborate, adaptive approaches. 

Extending on previous studies, we propose to treat cortices as dissipative thermodynamic systems that by homeostasis hold themselves near a critical level of activity that is far from equilibrium but steady state, a pseudo-equilibrium. We utilize FreemanÕs neuroscience insights manifested in the hierarchical brain model: the K (Katchalsky) sets \cite{katchalsky71, freeman75}. Originally, K-sets have been described mathematically using ordinary differential equations (ODE) with distributed parameters and with stochastic differential equations (SDE) using additive and multiplicative noise \cite{chang98noise, chang98homeo}. This approach has produced significant results, but certain shortcoming have been pointed out as well \cite{freeman1987, freeman1991, freeman1995}.  Calculating stable solutions for large matrices of nonlinear ODE and SDE that closely correspond to chaotic ECoG activity are prohibitively difficult and time-consuming to model on both digital and analog platforms \cite{principe01}. In addition, there are unsolved theoretical issues in constructing solid mathematics with which to bridge the difference across spatial and temporal scales between microscopic properties of single neurons and macroscopic properties of vast populations of neurons \cite{freeman1992, freeman1999}. 

In the past decade, neuropercolation approach has proved to be an efficient tool to address the above shortcomings by implementing K-sets using concepts of discrete mathematics and random graph theory \cite{npercscholar07}. Neuropercolation is a thermodynamics-based random cellular neural network model, which is closely related to cellular automata (CA), the field pioneered by Von Neumann who anticipated the significance of CA in the context of brain-like computing \cite{vonneumann66}. The present study is based on applying neuropercolation to systematically implement Freeman's principles of neurodynamics. Brain dynamics is viewed as a sequence of intermittent phase transitions in an open system with synchronization-desynchronization effects demonstrating symmetry breaking demarcated by spatio-temporal singularities \cite{freeman2004, kozmachaos08, fmascholar10, freemanqq12}. 

This work starts with the description of the basic building blocks of neurodynamics. Next, we develop a hierarchy on neuropercolation models with increasing complexity in structure and dynamics. Finally, we employ neuropercolation to describe critical behavior of brains and to interpret experimentally observed ECoG/EEG dynamics manifesting learning and higher cognitive functions. We conclude that criticality is a key aspect of the operation of brains and it is a basic attribute of intelligence in animals and in man-made devices.

\section{Principles of Hierarchical Brain Models}

\subsection{Freeman K Models: Structure and Functions}
 
We propose a hierarchical approach to spatio-temporal neurodynamics, based on K sets. Low-level K sets were introduced by Freeman in the 70Õs, named in the honor of Aharon Kachalsky, an early pioneer of neural dynamics \cite{freeman75}.  K sets are multi-scale models, describing increasing complexity of structure and dynamical behaviors. K sets are mesoscopic models, and represent an intermediate-level between microscopic neurons and macroscopic brain structures. The basic K0 set describes the dynamics of a cortical microcolums with about \(10^{4}\) neurons. K-sets are topological specifications of the hierarchy of connectivity in neuron populations. When first introduced, K-sets have been modeled using a system of nonlinear ordinary differential equations (ODE) \cite{freeman75, freeman1995}. K-dynamics predict the oscillatory waveforms that are generated by neural populations. K-sets describe the spatial patterns of phase and amplitude of the oscillations. They model observable fields of neural activity comprising EEG, LFP, and MEG.  K sets form a hierarchy for cell assemblies with the following components \cite{kscholar08}:
\begin{itemize}
\item
K0 sets represent non-interactive collections of neurons with globally common inputs and outputs: excitatory in K0e sets and inhibitory in K0i sets. The K0 set is the module for K-sets.
\item
KI sets are made of a pair of interacting K0 sets, both either excitatory or inhibitory in positive feedback. The interaction of K0e sets gives excitatory bias; that of K0i sets sharpens input signals.
\item
KII sets are made of a KIe set interacting with a KIi set in negative feedback giving oscillations in the gamma and high beta range (20-80 Hz). Examples include the olfactory bulb and the prepyriform cortex.
\item
KIII sets made up of multiple interacting KII sets. Examples include the olfactory system and the hippocampal system. These systems can learn representations and do match-mismatch processing exponentially fast by exploiting chaos.
\item
KIV sets made up of interacting KIII sets are used to model multi-sensory fusion and navigation by the limbic system.
\item
KV sets are proposed to model the scale-free dynamics of neocortex operating on and above KIV sets in mammalian cognition.
\end{itemize}

K sets are complex dynamical systems modeling the classification in various cortical areas, having typically hundreds or thousands of degrees of freedom. KIII sets have been applied to solve various classification and pattern recognition problems \cite{freeman1995, chang98, kozma2001a}. In early applications, KIII sets exhibited extreme sensitivity to model parameters which prevented their broad use in practice \cite{chang98}. In the past decade systematic analysis has identified regions of robust performance and stability properties of K-sets have been derived \cite{xu04, ilin2006}. Today, K sets are used in a wide range of applications, including detection of chemicals \cite{gutierrez03}, classification \cite{chang98, freeman01}, time series prediction \cite{beliaev07} and robot navigation \cite{harter05, kozmacim07}.  

\subsection{Basic Building Blocks of Neurodynamics} 
 
The hierarchical K model-based approach is summarized in the {\it 10 Building Blocks of Neurodynamics} \cite{freeman75, freeman1992, freeman1999}: 
\begin{enumerate}
 \item
 Non-zero point attractor generated by a state transition of an excitatory population starting from a point attractor with zero activity. This is the function of the KI set.  
 \item 
 Emergence of damped oscillation through negative feedback between excitatory and inhibitory neural populations. This is the feature that controls the beta-gamma carrier frequency range and it is achieved by KII having low feedback gain.  
 \item
 State transition from a point attractor to a limit cycle attractor that regulates steady state oscillation of a mixed E-I KII cortical population. It is achieved by KII with sufficiently high feedback gain. 
\item
The genesis of broad-spectral, aperiodic/chaotic oscillations as background activity by combined negative and positive feedback among several KII populations; achieved by coupling KII oscillators with incommensurate frequencies.  
\item
The distributed wave of chaotic dendritic activity that carries a spatial pattern of amplitude modulation AM in KIII. 
\item
The increase in nonlinear feedback gain that is driven by input to a mixed population, which results in the destabilization of the background activity and leads to emergence of an AM pattern in KIII as the first step in perception. 
\item
The embodiment of meaning in AM patterns of neural activity shaped by synaptic interactions that have been modified through learning in KIII layers. 
\item
Attenuation of microscopic sensory-driven noise and enhancement of macroscopic AM patterns carrying meaning by divergent-convergent cortical projections in KIV.  
\item
Gestalt formation and preafference in KIV through the convergence of external and internal sensory signals leading to the activation of the attractor landscapes leading to intentional action. 
\item
Global integration of frames at the theta rates through neocortical phase transitions representing high-level cognitive activity in the KV model. 
\end{enumerate}
 
Principles 1 through 7 describe the steps towards basic sensory processing, including pattern recognition, classification, and prediction, which is the function of KIII models. Principles 8 and 9 reflect the generation of basic intentionality using KIV sets. Principle 10 expresses the route to high-level intentionality and ultimately consciousness. The greatest challenge in modeling cortical dynamics is posed by the requirement to meet two seemingly irreconcilable requirements. One is to model the specificity of neural
action even to the level that a single neuron can be shown to have the possibility of capturing brain output. The other is to model the generality by which neural activity is synchronized and coordinated throughout the brain during intentional behavior. Various sensory cortices exhibit great similarity in their temporal dynamics, including the presence of spontaneous background activity, power-law distribution of spatial and temporal power spectral densities, repeated formation of AM spatial patterns with carrier frequencies in the beta and gamma ranges, and frame recurrence rates in the theta range. 

Models based on ODE and SDE have been used successfully to describe the mesoscopic dynamics of cortical populations for autonomous robotics \cite{kozmacim07, kozmasrr08}. However, such approaches suffered from inherent shortcomings. They falter in attempts to model the entire temporal and spatial range including the transitions between levels, which appear to take place very near to criticality. Neuropercolation and random graph theory offers a new approach to describe critical behavior and related phase transitions in brains. It is shown that criticality in the neuropil is characterized by a critical region instead of a singular critical point, and the trajectory of the brain as a dynamical systems crosses the critical region from less organized (gaseous) phase to more organized (liquid) phase during input induced destabilization and vise versa \cite{freemanviti08, freeman12, freeman12nn}. Neuropercolation is able to simulate these important results from mesoscopic ECOG/EEG recording across large spacial and temporal scales, as will be introduced in this essay.

\subsection{Motivation of Neuropercolation Approach to Neurodynamics}

We utilize the powerful tools of random graph theory (RGT) developed over the past 50 years \cite{erdos1959, bollobas1984, bollobas2006} to establish rigorous models of brain networks. Our model incorporates cellular automata and percolation theory in random graphs that are structured in accordance with cortical architectures. We use the hierarchy of interactive populations in networks as developed in Freeman K models \cite{freeman75}, but replace differential equations with probability distributions from the observed random graph that evolve in time. The corresponding mathematical object is called neuropercolation \cite{npercscholar07}. Neuropercolation theory provides a suitable mathematical approach to describe phase transitions and critical phenomena in large-scale networks. It has potential advantage compared to ODEs and PDEs when modeling spatio-temporal transitions in the cortex, as differential equations assume some degree of smoothness in the described phenomena, which may not be very suitable to model the observed sudden changes in neurodynamics. Neural percolation model is a natural mathematical domain for modeling collective properties of brain networks, especially near critical states, when the behavior of the system changes abruptly with the variation of some control parameters. 

Neuropercolation considers populations of cortical neurons which sustain their metastable state by mutual excitation, and that its stability is guaranteed by the neural refractory periods. Nuclear physicists have used the concept of criticality to denote the threshold for ignition of a sustained nuclear chain reaction, i.e., fission. The critical state of nuclear chain reaction is achieved by a delicate balance between the material composition of the reactor and its geometrical properties. The criticality condition is expressed as the identity of geometrical curvature (buckling) and material curvature. Chain reactions in nuclear processes are designed to satisfy strong linear operational regime conditions, in order to assure stability of the underlying chain reaction. That usage fails to include the self-regulatory processes in systems with nonlinear homeostatic feedback that characterize cerebral cortices. 

A key question is how cortex transits between gas-like randomness and liquid-like order near the critical state. We have developed thermodynamic models of cortex  \cite{freeman12, freeman12nn}, which postulates two phases of neural activity: vapor-like and liquid-like. In the vapor-like phase the neurons are uncoupled and maximally individuated, which is the optimal condition for processing microscopic sensory information at low density. In the liquid-like phase the neurons are strongly coupled and thereby locked into briefly stable macroscopic activity patterns at high density, such that every neuron transmits to and receives from all other neurons by virtue of small-world effects \cite{watts98, sporns06}. Local 1/f fluctuations have the form of PM patterns that resemble droplets in vapor. Large-scale, spatially coherent AM patterns emerge from and dissolve into this random background activity but only on receiving a CS. They do so by spontaneous symmetry breaking \cite{freemanviti08} in a phase transition that resembles condensation of a rain drop, in that it requires a large distribution of components, a source of transition energy, a singularity in the dynamics, and a connectivity that can sustain interaction over relatively immense correlation distances. We conclude that the background activity at the pseudo-equilibrium state conforms to self-organized criticality, that the fractal distribution of phase patterns corresponds to that of avalanches, that the formation of a perception from sensory input is by a phase transition from a gas-like, disorganized, low-density phase to a liquid-like high-density phase \cite{freeman12}. 

\section{Mathematical Formulation of Neuropercolation}

\subsection{Random Cellular Automata on a Lattice} 

In large networks, such as cortex, organized dynamics emerges from the noisy and chaotic behavior of a large number of microscopic components. Such systems can be modeled as graphs, in which neurons become vertices. The activity of every vertex evolves in time depending on its own state, the states of its neighbors, and possibly some random influence. This leads us to the general formulation of random cellular automata. In a basic two-state cellular automaton, the state of any lattice point \(x \in \mathbb{Z}^d\) is either active or inactive. The lattice is initialized with some (deterministic or random) configuration. The states of the lattice points are updated (usually synchronously) based on some (deterministic or probabilistic) rule that depends on the activations of their neighborhood. For related general concepts, see cellular automata such as Conway's Game of Life, cellular neural network, as well as thermodynamic systems like the Ising model \cite{berlekamp82, cipra1987, marcq1997, wolfram02}.
Neuropercolation models develop neurobiologically motivated generalizations of cellular automata models and incorporate the following major conditions: 


\begin{itemize}
\item
{\it Interaction with noise}: The dynamics of the interacting neural populations is inherently non-deterministic due to dendritic noise and other random effects in the nervous tissue and external noise acting on the population. Randomness plays a crucial role in neuropercolation models and there is an intimate relationship between noise and the system dynamics, due to the excitable nature of the neuropil.
\item
{\it Long axon effects (rewiring)}: Neural populations stem ontogenetically in embryos from aggregates of neurons that grow axons and dendrites and form synaptic connections of steadily increasing density.  At some threshold the density allows neurons to transmit more pulses than they receive, so that an aggregate undergoes a state transition from a zero point attractor to a non-zero point attractor, thereby becoming a population. In neural populations, most of the connections are short, but there are a relatively few long-range connections mediated by long axons. The effect of long-range axons are similar to small-world phenomena \cite{watts1998} and it is part of the neuropercolation model.
\item
{\it Inhibition}: An important property of neural tissue that it contains two basic types of interactions: excitatory and inhibitory ones. Increased activity of excitatory populations positively influence (excite) their neighbors, while highly active inhibitory neurons negatively influence(inhibit) the neurons they interact with. Inhibition is inherent in cortical tissues and it controls stability and metastability observed in brain behaviors \cite{jaskelso95, ilin2006}. Inhibitory effects are are part of neuropercolation models.
\end{itemize}

\subsection{Update Rules}

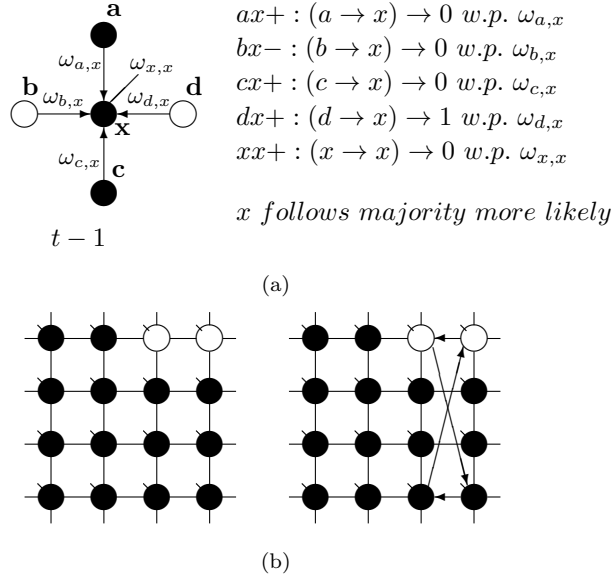
\begin{figure}[ht!]
\begin{center}
\subfigure[]{ \label{majorityEG} \input{majorityRuleS.tex} } 
\subfigure[]{ \label{subgraphEG}\input{subgraphFig.tex} }
\caption{\label{majorityEGsubgraphEG}
Illustration of update rules in 2D lattice. (a): An example of how majority function works. \(bx\) edge inhibits with strength \(\omega_{b,x}\), so it sends 0 when \(s(b,t-1)=1\). Given the scenario, \(s(x,t)\) is 0 most likely.
\ref{subgraphEG}: Example of 2-D torus of order \(4\times4\). First row/column is connected with last row/column. Each vertex has a self-influence. Right: 2-D torus after the basic random rewiring strategy. Two out of eighty (\(16\times5\)) edges are rewired or 2.5\%.
}
\end{center}
\end{figure}

A vertex \(v_{i}\in V\) of a graph \(G(V,E)\) is in one of two states, \(s(v_{i})\), and influenced via the edges by \(d(v_{i})\) neighbors. An edge from \(v_{i}\) to \(v_{j}\), \(v_{i}v_{j}\in E\), either excites and inhibits. Excitatory edges project the states of neighbors and inhibitory edges project the opposite states of neighbors, 0 if the neighbor's state is 1, and 1 if it is 0.
Vertex's state, influenced by edges, is determined by the majority rule; when the most neighbors are active, the higher a chance for the vertex to be active, and when the most neighbors are inactive, the higher a chance for the vertex to be inactive.
At time \(t=0\) \(s(v_{i})\) is randomly set to \(0\) or \(1\). Then, for \(t=1,2,...,T-1\), a majority rule is applied simultaneously over all vertices. A vertex \(v_{i}\) is influenced by a state of its neighbor \(v_{j}\in N(v_{i})\), whenever a random variable \(R(v_{i},t)\) is less than the influencing excitatory edge \(v_{j}v_{i}\) strength, \(\omega_{j,i}\), else the vertex \(v_{i}\) is influenced by an opposite state of neighbor \(v_{j}\).
If the edge \(v_{j}v_{i}\) is inhibitory, the vertex \(v_{j}\) sends 0 when \(s(v_{j})=1\), and 1 when \(s(v_{j})=0\).
Then, a vertex \(v_{i}\) gets a state of the most common influence, if there is such, otherwise a vertex state is randomly set to 0 or 1, Fig. \ref{majorityEG}.

Formula for the majority rule:
\begin{align}
s(v_{i},t)\!&=\!\left\{
\begin{array}{l}
\mbox{0 if}\displaystyle\sum_{\forall v_{j}\in N(v_{i})}^{}f(v_{j},t) < \frac{d(v_{i})}{2} \\ \\
\mbox{1 if}\displaystyle\sum_{\forall v_{j}\in N(v_{i})}^{}f(v_{j},t) > \frac{d(v_{i})}{2} \\ \\
\mbox{0 or 1 if}\displaystyle\sum_{\forall v_{j}\in N(v_{i})}^{}f(v_{j},t) = \frac{d(v_{i})}{2}
\end{array}
\right.
{\label{majorityRule}}
\end{align}
\begin{align}
f(v_{j},t)&=\!\left\{
\begin{array}{l}
\!0\mbox{ if }\omega_{j,i}\bigl(s(v_{j},t\!-\!1)\!=\!0\bigr)\!\ge\!R(v_{i},t)\mbox{; else } 1 \\
\!1\mbox{ if }\omega_{j,i}\bigl(s(v_{j},t\!-\!1)\!=\!1\bigr)\!\ge\!R(v_{i},t)\mbox{; else } 0
\end{array}
\right. \notag
\end{align}
\begin{align}
s(v_{j},t\!-\!1)&=\!\left\{
\begin{array}{l}
0\mbox{ if }\omega_{j,i}\bigl(s(v_{j},t\!-\!1)\!=\!0\bigr)\mbox{ excites; else } 1 \\
1\mbox{ if }\omega_{j,i}\bigl(s(v_{j},t\!-\!1)\!=\!1\bigr)\mbox{ excites; else } 0
\end{array}
\right. \notag
\end{align}
\begin{align}
0&\le R(v_{i},t)\le1 \notag \\
0.5&\le \omega_{j,i}\bigl(s(v_{j})\bigr)\le1 \notag
\end{align}

\subsection{Two-Dimensional Lattice with Rewiring}

Lattice-like graphs are built by randomly re-connecting some edges in the graphs set on a regular 2-dimensional grid in and folded into tori. In the applied basic random rewiring strategy, \(n\) directed edges from \(n\times2\) vertices are plugged out from a graph randomly. At that point, the graph has \(n\) vertices lacking an incoming edge and \(n\) vertices lacking an outgoing edge. To preserve the vertex degrees of the original graph, the set of plugged-out edges is returned back to the vertices with the missing edges in the random order. An edge is pointed to the vertex missing the incoming edge and is projected from the vertex missing the outgoing edge, Fig. \ref{subgraphEG}. Such a two-dimensional graph models a KI sets with homogeneous population of excitatory or inhibitory neural population.

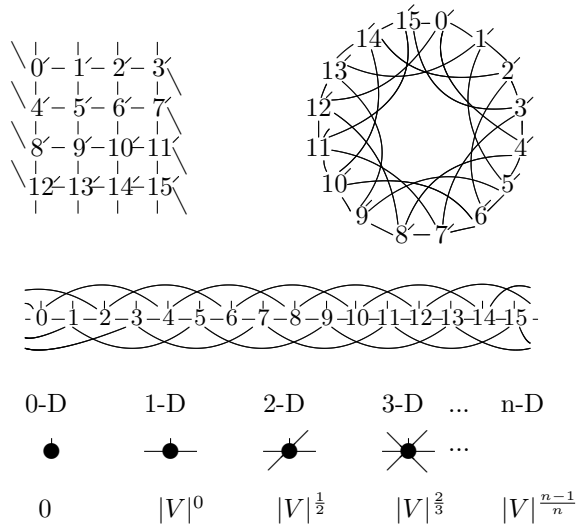
\begin{figure}[ht!]
\begin{center}
\input{nDapprox.tex}\\
\input{modelStructure.tex}\\
\input{generalModel.tex}\\
\caption{\label{modelStructure}
Top-left: Example of labeled \(4\times4\) torus in 2-dimension. Top-right: Conversion of the 2D lattice approximately into an 2-D torus where vertices are set in circle. Middle: \(4\times4\) 2-D torus cut an unfolded into a linear band. Bottom section: illustration of the neighborhoods in higher dimensions. 
}
\end{center}
\end{figure}

It will be useful to reformulate the basic subgraph on a regular grid, as shown in Fig. \ref{modelStructure}. 
We list all vertices along a circle and connect the corresponding vertices to obtain the circular representation on the top right corner of Fig. \ref{modelStructure}. This circle is unfolded into a linear band by cutting it up at one location, as shown in the middle of the figure. Such representation is very convenient for the design of the rewiring algorithm. Note that this circular representation is not completely identical to the original lattice, due to the different edges when folding the lattice into torus. We have conducted experiments to compare the two representations and obtained that the difference is minor and do not affect the basic conclusions. An important advantage of this band representation that it can be generalized to higher dimensions by simply adding suitable new edges between the vertices. For example, in 2D we have 4 direct neighbors and thus 4 connecting edge, in 3D we have 6 connecting edges, and so on. generalization to higher dimensions is not used in this work as the 2D model is a reasonable approximation of the layered cortex.

\subsection{Double-Layered Lattice}
We construct a double-layered graph by connecting two lattices, see Fig. \ref{oscillatorEG2}. The top layer  \(G_{0}\) is excitatory, while the bottom layer  \(G_{1}\) is inhibitory. The excitatory subgraph \(G_{0}\) projects to the inhibitory subgraph \(G_{1}\) through edges that excite, while \(G_{1}\) inhibitory layer projects toward \(G_{0}\) with edges that inhibit. 
An excitatory edge from \(G_{0}\) influence the vertex of \(G_{1}\) with 1 when active and with 0 when inactive. Conversely, an inhibitory edge from \(G_{1}\) influences the vertex of \(G_{0}\) with 0 when active and with 1 when inactive.

\subsection{Coupling Two Double-Layered Lattices}
\begin{figure}[ht!]
\begin{center}
\subfigure[]{ \label{oscillatorEG2} \input{oscillatorEG2.tex} } \\
\subfigure[]{ \label{oscillator2EG2} \input{oscillator2EG2.tex} }
\caption{\label{oscillatorEG2oscillator2EG2}
 Illustration of coupled layers with randomly selected connection weights between them. 
\ref{oscillatorEG2}:Example of a double-layer coupled system that includes an excitatory layer \(G_{0}\) and an inhibitory layer \(G_{1}\). Within the layers the edges are excitatory; edges from \(G_{0}\) to \(G_{1}\) are excitatory and edges from \(G_{1}\) to \(G_{0}\) are inhibitory.
\ref{oscillator2EG2}: Example of two coupled double-layers, all together 4 coupled layers: \(G_{0}\) and \(G_{2}\) are excitatory layers, while \(G{1}\) and \(G_{3}\) denote inhibitory layers. Each subgraph has three edges rewired.
Edges from \(G_{1}\) to \(G_{0}\) and edges from \(G_{3}\) to \(G_{2}\) are inhibitory; the other edges are excitatory.
}
\end{center}
\end{figure}
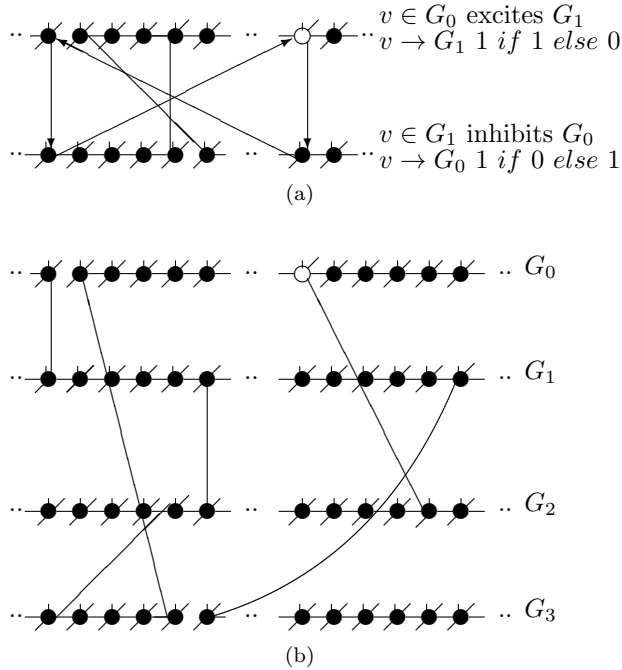

By coupling two double-layered graphs, we have four subgraphs interconnected to form a graph \(G=G_{0}\cup G_{1}\cup G_{2}\cup G_{3}\). Subgraph \(G_{0}\) and \(G_{1}\) are coupled into one double layer and \(G{2}\) and \(G_{3}\) into the other.
There are 3 excitatory edges between \(G_{0}\) and \(G_{2}\), \(G_{0}\) and \(G_{3}\), \(G_{1}\) and \(G_{2}\), \(G_{1}\) and \(G_{3}\), respectively, which are responsible to couple the two double-layered lattices, Fig. \ref{oscillator2EG2}.

\subsection{Statistical Characterization of Critical Dynamics of Cellular Automata}

The most fundamental parameter describing the state of a finite graph \(G\) at time \(t\) is its overall activation level \(S(G,t)\) defined as follows:
\begin{align}
S(G,t)&=\displaystyle\sum_{i=0}^{\vert V\vert-1}s(v_{i}) {\label{activation}}
\end{align}

It is useful to introduce the normalized activation per vertex, \(a(G,t)\), which is zero if all sites are inactive and 1 if all sites are active. In general, \(a(G,t)\) is between 0 and 1:
\begin{align}
a(G,t)&=\frac{S(G,t)}{\vert V\vert} {\label{density}} 
\end{align}

The average normalized activation over time interval \(T\) is given as:
\begin{align}
\langle a\rangle&=\frac{\displaystyle\sum_{t=0}^{T-1}a(G,t)}{T} \\  
\langle a^{*}\rangle&=\frac{\displaystyle\sum_{t=0}^{T-1}\vert a(G,t)-0.5\vert}{T}
\end{align}

The latter quantity can be interpreted as the average mangetization in terms of statistical physics. Depending on their structure, connectivity, update rules, and initial conditions, probabilistic automata can display very complex behavior as they evolve in time. Their dynamics may converge to a fixed point, to limit cycles, or to chaotic attractors. In some specific models, it is rigorously proven that the system is bistable and exhibits critical behavior. Namely, it spends a long time in either low- or high-density configurations, with mostly 0 or 1 activation values at the nodes, before a very rapid transition to the other state \cite{balister2006, kozma2005}. In some cases, the existence of phase transitions can be shown mathematically. For example, the magnitude of the probabilistic component \(\omega\) of the random majority update rule in Fig. \ref{majorityEG} is shown to have  a critical value, which separates the region with stable fixed point from the bistable regime \cite{npercscholar07}. 

In the lack of exact mathematical proofs, Monte Carlo computer simulations and methods of statistical physics provide us with detailed information on the system dynamics. The critical state of the system is studied using the statistical moments. Based on the finite size scaling theory, there are statistical measures which are invariant to system size at the state of criticality \cite{binder1981}. Evaluating these measures, the critical state can be found. It has been shown that in addition to the system noise, which acts as a temperature measure in the model, there are  other suitable control parameters which can drive the system to critical regimes, including the extent of rewiring and the inhibition strength. \cite{puljic2008}. 

Denote this quantity \(\omega\) which will be a control parameter; it can be the noise, rewiring, connection strength or other suitable quantity.  In this work, \(\omega\) describes noise effects and related to the system noise level \(\epsilon = 1 - \omega\). 
Distributions of \(a(G,\omega)\) and \(a^{*}(G,\omega)\) depend on \(\omega\). At a critical value of \(\omega_{i}\), the probability distribution functions of these quantities suddenly change.
\(\omega_{i}\) values are traditionally determined by evaluating the 4th order moments (kurtosis, peakedness measure) \(u_{4}\), \cite{binder1981,ferrenberg1988,li1995,loison1999,hasenbusch2008}, but they can also be measured using the 3rd order moments (skewness measure) or \(u_{3}^{*}\):
\begin{align}
u_{4}(G)&=\frac{\langle(a-\langle a\rangle)^{4}\rangle}{\langle(a-\langle a\rangle)^{2}\rangle^2} {\label{kurtosis}} \\
u_{3}^{*}(G)&=\frac{\langle(a^{*}-\langle a^{*}\rangle)^{3}\rangle}{\langle(a^{*}-\langle a^{*}\rangle)^{2}\rangle^{3/2}} {\label{skwenessA}}
\end{align}
According to finite-size scaling theory, for two structurally equivalent finite graphs \(G\) and \(G'\) of different sizes,
\(u_{4}(G)\ne u_{4}(G')\) and \(u_{3}^{*}(G)\ne u_{3}^{*}(G')\) for \(\omega\ne\omega_{i}\). However, for \(\omega=\omega_{i}\), 
\(u_{4}(G)=u_{4}(G')\) and \(u_{3}^{*}(G)=u_{3}^{*}(G')\). In this study, we use both the 3rd and 4th order momentums to determine the critical points of the system.

\section{Critical Regimes of Coupled Hierarchical Lattices}

\subsection{Dynamical Behavior of 2-D Lattices with Rewiring}

Two-dimensional lattices folded into tori have been thoroughly analyzed in \cite{kozma2005}; the main results are summarized here. In the absence of rewiring, the lattice is in one of the two possible states: ferromagnetic and paramagnetic, depending on the noise strength \(\omega\). The following states are observed:
\begin{itemize}
\item
For \(\omega<\omega_{0}\), states 0 and 1 are equiprobable across the graph and \(a\) distribution is uni-modal. This state corresponds to paramagnetic regime. 
\item
For \(\omega>\omega_{0}\), one state dominates; vertex states are mostly 0 or mostly 1, and graph's \(a\) distribution is bi-modal. This is the ferromagnetic condition.
\item
In the close neighborhood of the critical point \(\omega_{0}\), the lattice dynamics is unstable. Immediately above \(\omega_{0}\), the vertex mostly follows its majority influences.
\end{itemize}

\begin{figure}[ht!]
\begin{center}
\label{egBinderMix3b} \includegraphics[width=0.9\textwidth, height=0.5\textwidth]{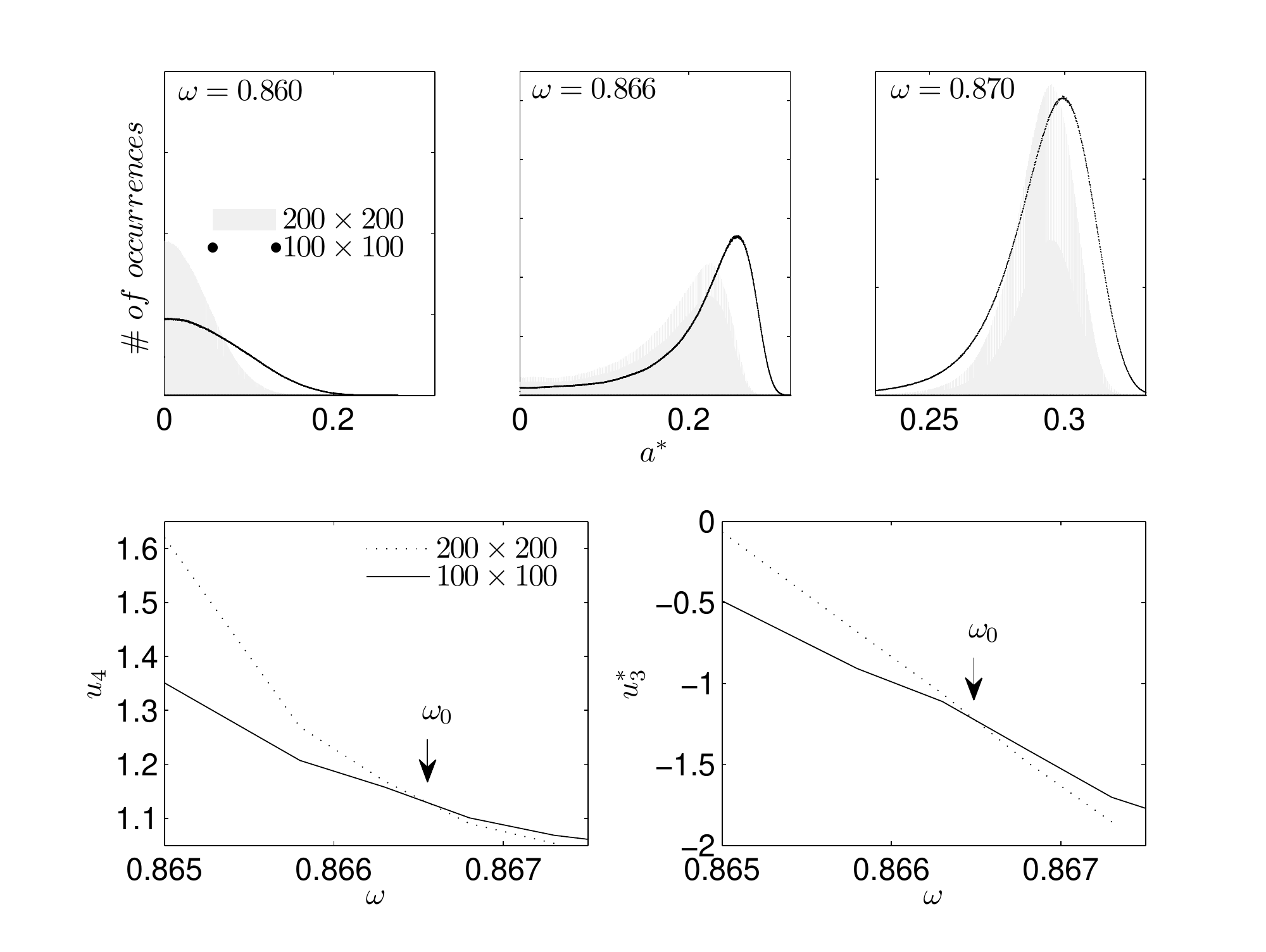} 
\caption{\label{egBinderMix3b}
Illustration of the dynamics of the 2-D torus; the case of KI architecture.
There are two distinct dynamic regimes; top left diagram: unimodal pdf with zero average activation (paramagnetic regime); top middle and right: the case of bimodal pdf with a positive activation level(ferromagnetic regime).
Lower part shows \(u_{4}=\langle(a-\langle a\rangle)^{4}\rangle/\langle(a-\langle a\rangle)^{2}\rangle^2\) (left) and \(u_{3}^{*}\) around \(\omega_{0}\) for 2 graphs of different orders.
}
\end{center}
\end{figure}

\begin{figure}
\begin{center}
\includegraphics[width=0.9\textwidth, height=0.5\textwidth]{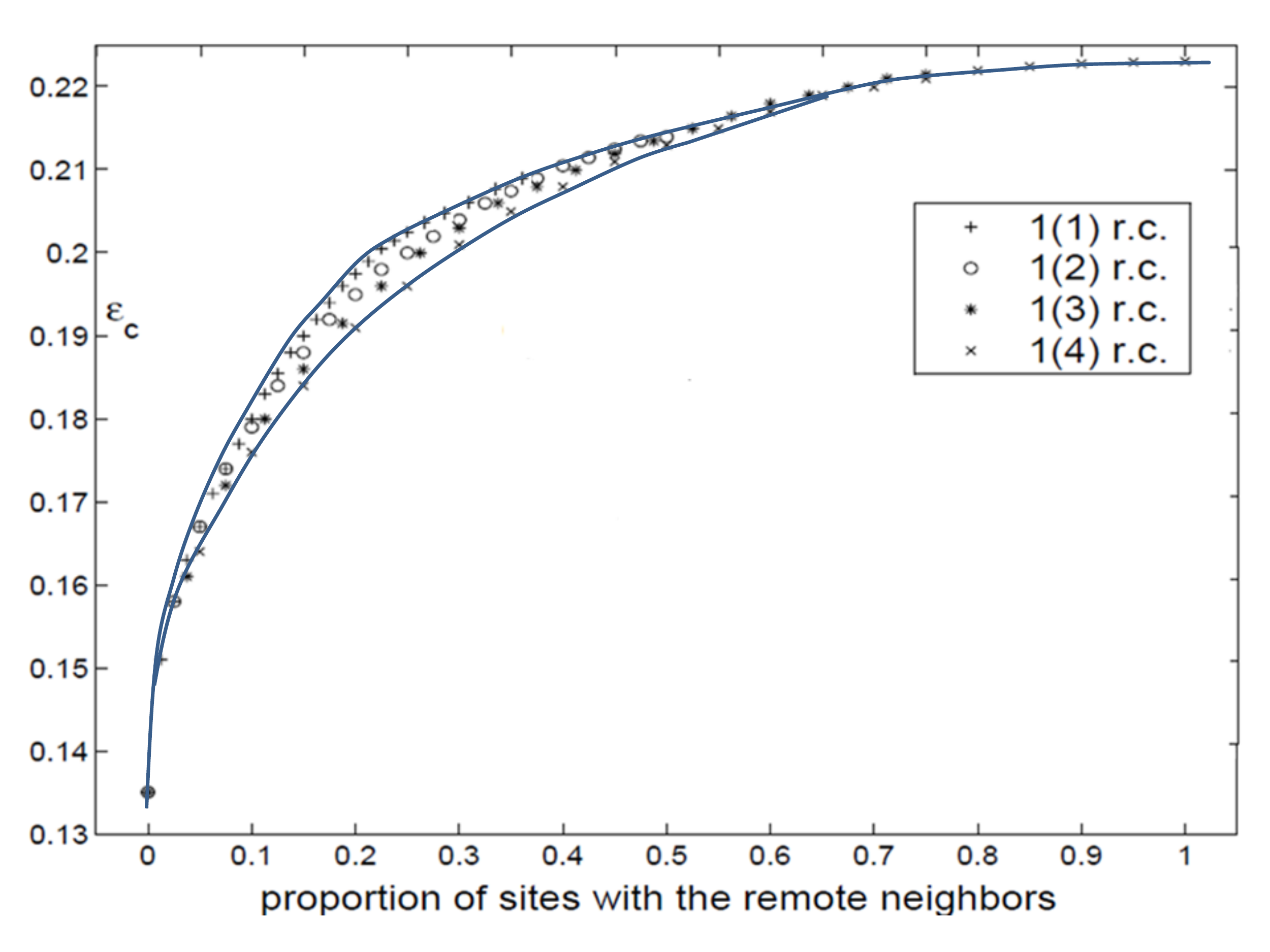}
\caption{\label{banana3}
Integrated view of the relationship between the proportion of rewired edges and critical noise \(\epsilon_{c} = 1 -\omega{o}\), adopted from \cite{kozma2005}. Notations 1(i), 1(2), 1(3), and 1(4) mean that a vertex can have 1, 2, 3, or 4 remote connections, respectively.
}
\end{center}
\end{figure}

For \( \omega > \omega_{0}\), the vertices cease to act individually and become a part of a group. Their activity level is determined by the population. This regime is defined by the condition that the vertices project more common activation than their receive, in the probabilistic sense. This transition is the first building block of neurodynamics and it corresponds to KI level dynamics. Fig. \ref{egBinderMix3b} illustrates critical behavior of a 2-D torus with \(\omega_{0}\approx0.866\); no rewiring. Calculations show that for the same torus having 5.00\% edges rewired, the critical probability changes to \(\omega_{0}\approx0.830\). 
Rewiring leads to a behavior of paramount importance for future discussions: the critical condition is manifested as a {\it critical region}, instead of a singular {\it critical point}. We will show various manifestations of such principle in the brain. In the case of rewiring discussed in this section, the emergence of critical region is related to the fact that same amount of rewiring can be achieved in multiple ways. For example, it is possible to rewire all edges with equal probability, or select some vertices which are more likely rewired etc. In real life such as brains, all these conditions happen simultaneously. This behavior is illustrated in Fig. \ref{banana3} and it has been speculated as a key condition guiding the evolution of the neuropil in the infant's brain in the months following birth \cite{kozma2004, kozma2005, balister2006} giving rise to a robust, broad-band background activity as  a necessary condition of the formation of complex spatio-temporal patterns during cognition. 

\subsection{Narrow Band Oscillations in Coupled Excitatory-Inhibitory Lattices}

\begin{figure}[ht!]
\begin{center}
\includegraphics[width=0.9\textwidth, height=0.5\textwidth]{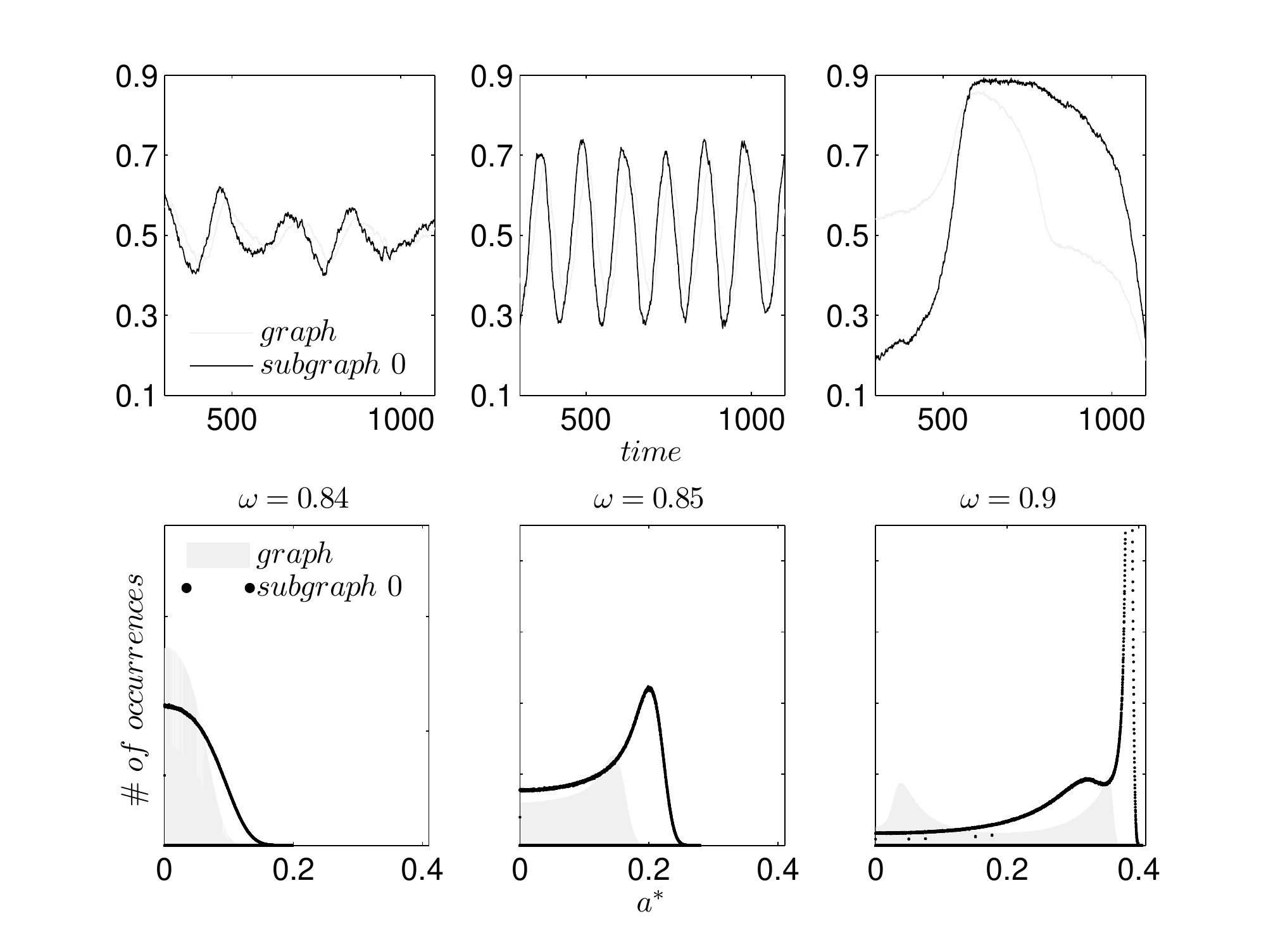}
\caption{\label{egBinderMix3b2_96_2304_1728_distributionMix0}
Illustration of the dynamics of coupled excitatory-inhibitory layers; the case of KII architecture. There are three modes of the oscillator \(G(5\%,3.75\%)\) made of two 2-D layers \(G_{0}\) and \(G_{1}\) of order \(96\times96\). 
Top: \(a(G,\omega)\) of the whole graph (gray line) and of subgraph \(G_{0}\) (black line) at three different \(\omega\) values, which produce three different dynamic modes. 
Bottom part shows distributions of \(a^{*}\) at 3 different \(\omega\) values for graphs of orders \(100\times100\) and \(200\times200\); only one side of the symmetric \(a^{*}\)distribution is shown. 
}
\end{center}
\end{figure}

Inhibitory connections in coupled subgraph  \(G=G_{0}\cup G_{1}\) can generate oscillations and multi-modal activation states \cite{puljic2008}. In an oscillator, there are three possible regimes demarcated by two critical points \(\omega_{0}\) and \(\omega_{1}\):
\begin{itemize}
\item
\(\omega_{0}\) is the transition point where narrow-band oscillations start.
\item
\(\omega_{1}\) is the transition point where narrow-band oscillations terminate.
\end{itemize}
Under appropriately selected graph parameters, \(a(G,\omega)\) distribution is uni-modal when \(\omega<\omega_{0}\), bi-modal when \(\omega_{0}<\omega<\omega_{1}\), and quadro-modal when \(\omega>\omega_{1}\), Fig. \ref{egBinderMix3b2_96_2304_1728_distributionMix0}. Each subgraph has 2304 (5\%) edges rewired within the layer and 1728 (3.75\%) edges rewired towards the other subgraph. 
Just below \(\omega_{0}\), vertices may oscillate if they are excited. \(a(G,\omega)\) values of temporarily excited graph oscillate, but return to the basal level in the absence of excitation. This form of oscillation is the second building block of neuron inspired dynamics.
When \(\omega\) is further increased, i.e., \(\omega_{0}<\omega<\omega_{1}\), \(a(G,\omega)\) exhibits sustained oscillations even in the absence of external perturbation. When the double-layer is temporarily excited, it returns to the basal oscillatory behavior. Self-sustained oscillation is the third building block of neuron inspired dynamics and it corresponds to KII hierarchy.

\section{Broad-Band Chaotic Oscillations}

\subsection{Dynamics of Two Double Arrays}

\begin{figure}[ht!]
\begin{center}
\subfigure[]{ \label{2ndAND3rdBBNDeg} \includegraphics[width=0.9\textwidth, height=0.5\textwidth]{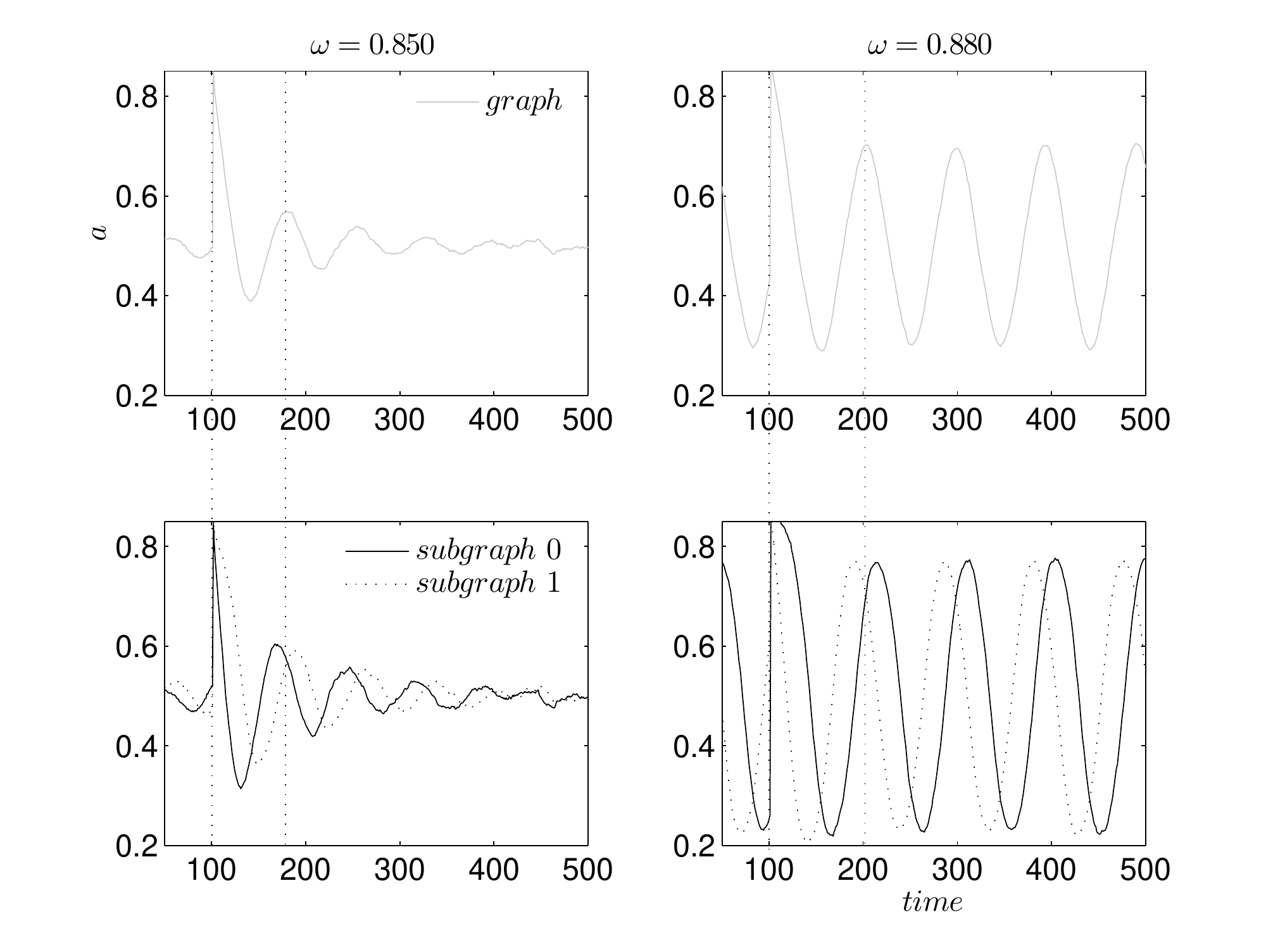} }
\subfigure[]{ \label{oscillationsMix3} \includegraphics[width=0.9\textwidth, height=0.5\textwidth]{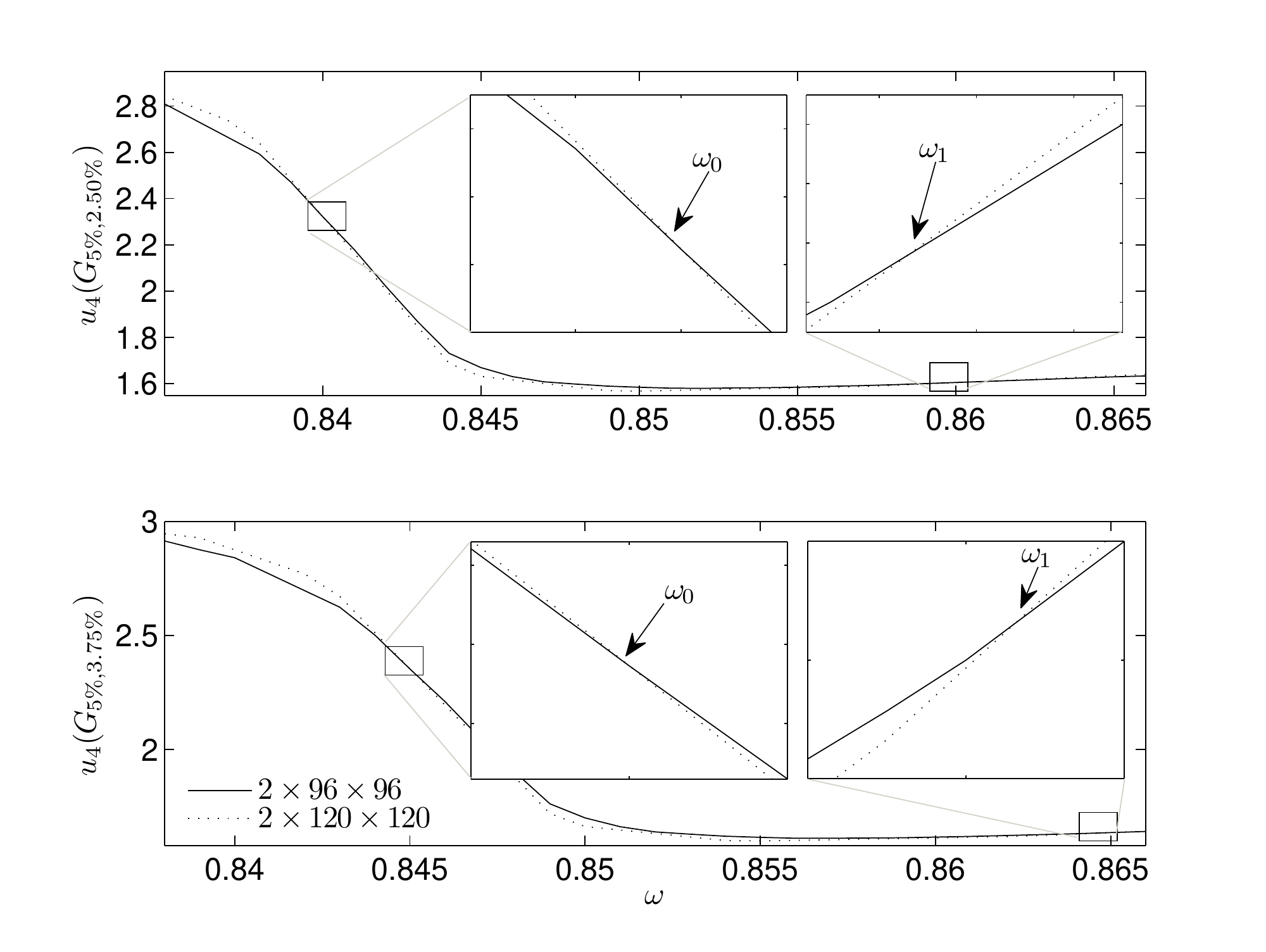} }
\caption{\label{2ndAND3rdBBNDegoscillationsMix3}
Critical states obtained by coupling two double layers of oscillators with rewiring (case of KIII).
(a): Left: the graph starts to oscillate after an impulse, but the oscillation decays and returns to the basal level; right: oscillation is sustained even after the input is removed, if \(\omega_{0}<\omega<\omega_{1}\).
(b): Critical behavior and \(\omega_{0}\) and \(\omega_{1}\) values for two different oscillators, \(G(5\%,2.5\%)\) (up) and \(G(5\%,3.75\%\) (bottom). 
}
\end{center}
\end{figure}

\begin{figure}[ht!]
\begin{center}
\includegraphics[width=0.9\textwidth, height=0.5\textwidth]{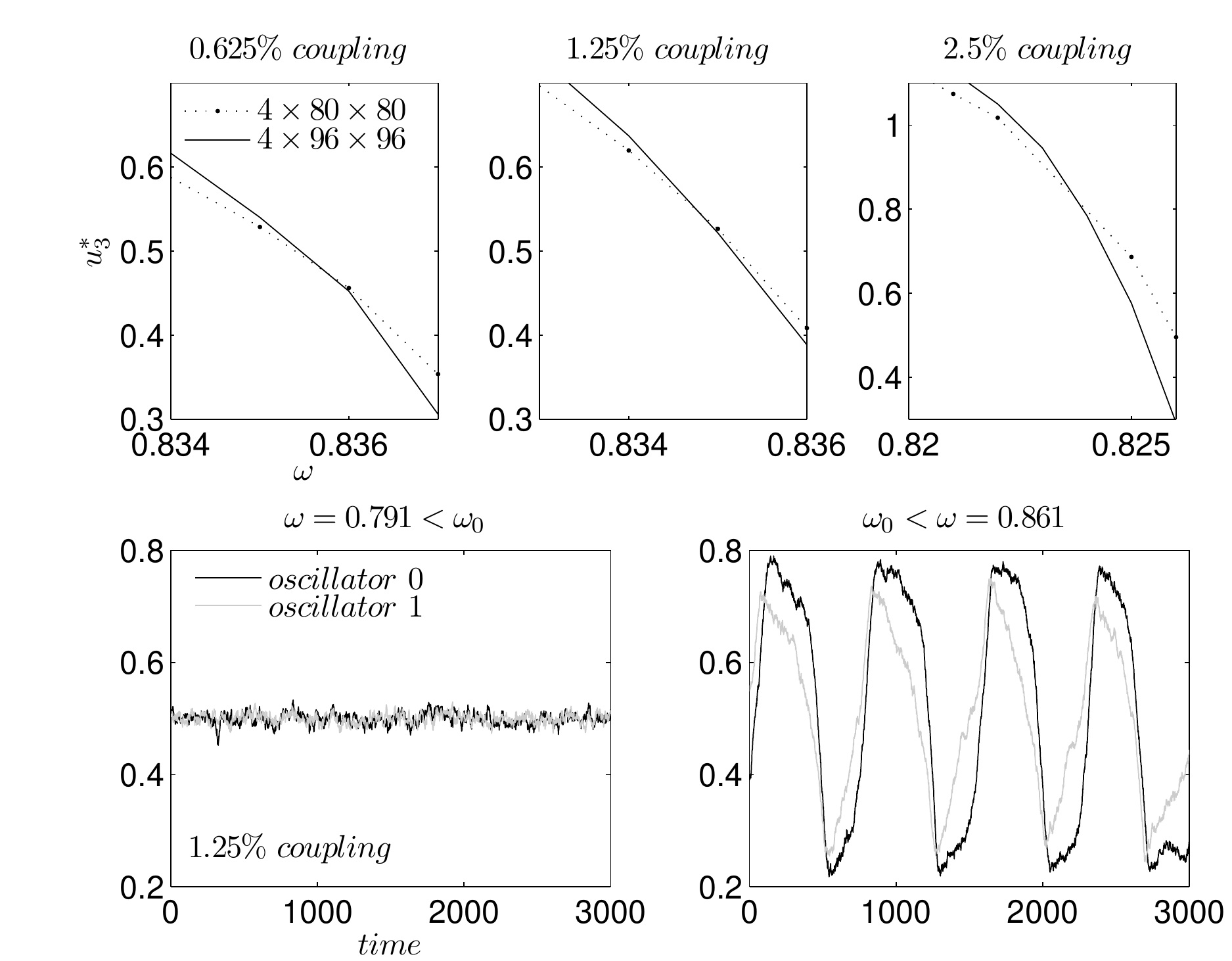}
\caption{\label{skewnessEG1}
Coupling two KII oscillators into KIII, lowers \(\omega_{0}\), the point above which large-scale synchronous oscillations are formed. 
The top plots show \(u_{3}^{*}\) values for \(G(5\%,2.5\%,3.75\%,0.625\%)\), \(G(5\%,2.5\%,\) \(3.75\%,1.25\%)\), and \(G(5\%,2.5\%,3.75\%,2.5\%)\).
Lower part shows typical activation of graph for \(\omega\) below and above \(\omega_{0}\).
}
\end{center}
\end{figure}

Two coupled layers with inhibitory and excitatory nodes exhibit narrow-band activity oscillations. These coupled lattices with appropriate topology can model any frequency range with oscillations occurring between \(\omega_{0}\) and \(\omega_{1}\). 
For example, an oscillator \(G(5\%,2.5\%)=G_{0}\cup G_{1}\) has 2.5\% edges rewired across the two layers and 5\% edges are rewired within each layer. \(G(5\%,2.5\%)\) has \(\omega_{0}\approx0.840\) and \(\omega_{1}\approx0.860\), Fig. \ref{oscillationsMix3}. Similarly, oscillator \(G(5\%,3.75\%)\) has \(\omega_{0}\approx0.845\) and \(\omega_{1}\approx0.865\). Each of the coupled lattices in this example is of size \(100 \times 100\).
When coupling two such oscillators together, we have a total of 4 layers (tori) coupled into a unified oscillatory system. Clearly, the two oscillators coupled together may or may not agree on a common frequency. We will show that under specific conditions, various operational regimes may occur, including narrow band oscillations, broad-band (chaotic) oscillations, and intermittent oscillations, and various combinations of these.
One of multiple ways to couple two oscillators is to rewire excitatory edges between subgraphs \(G_{0}\)-\(G_{2}\), \(G_{0}\)-\(G_{3}\), \(G_{1}\)-\(G_{2}\), and \(G_{1}\)-\(G_{3}\), respectively. Two coupled oscillators have additional behavioral regimes. When oscillators \(G(5\%,2.5\%)\) and \(G(5\%,3.75\%)\) are coupled with 0.625\% edges into a new graph \(G(5\%,  2.5\%,\). \(  3.75\%, 0.625\%)\), \(\omega_{0}\) is shifted to lower level, Fig. \ref{skewnessEG1}.
Every parameter describing a graph influences the critical behavior and under appropriate conditions. If two connected oscillators with different frequencies cannot agree on a common mode, but together they can generate large-scale synchronization or chaotic background activity.

In the graph made of two oscillators, there are four critical points \(\omega_{0}<\omega_{1}<\omega_{2}<\omega_{3}\),  separating 5 possible major operational regimes. 
\begin{itemize}
\item
Case of \(\omega<\omega_{0}\): The aggregate activation distribution is uni-modal and it represents the paramagnetic regime in lattice dynamics analogy.
\item
\(\omega_{0}<\omega<\omega_{1}\): The two coupled oscillators exhibit large-scale synchronization and the activation distribution is bi-modal.
\item
\(\omega_{1}<\omega<\omega_{2}\): The two coupled oscillators with different frequencies cannot agree on a common mode, so together they generate aperiodic background activity.
\item
\(\omega_{2}<\omega<\omega_{3}\): only one oscillator oscillates in a narrow band and the activation distribution is okta-modal.
\item
\(\omega>\omega_{3}\): none of the components exhibit narrow-band oscillation and the activation distribution is a hexa-modal. This corresponds to ferromagnetic state.
\end{itemize}
The above regimes correspond to the fourth building block of neurodynamics as described by KIII sets.

\begin{figure}[ht!]
\begin{center}
\subfigure[]{ \label{phaseLagsNS} \includegraphics[width=0.9\textwidth, height=0.5\textwidth]{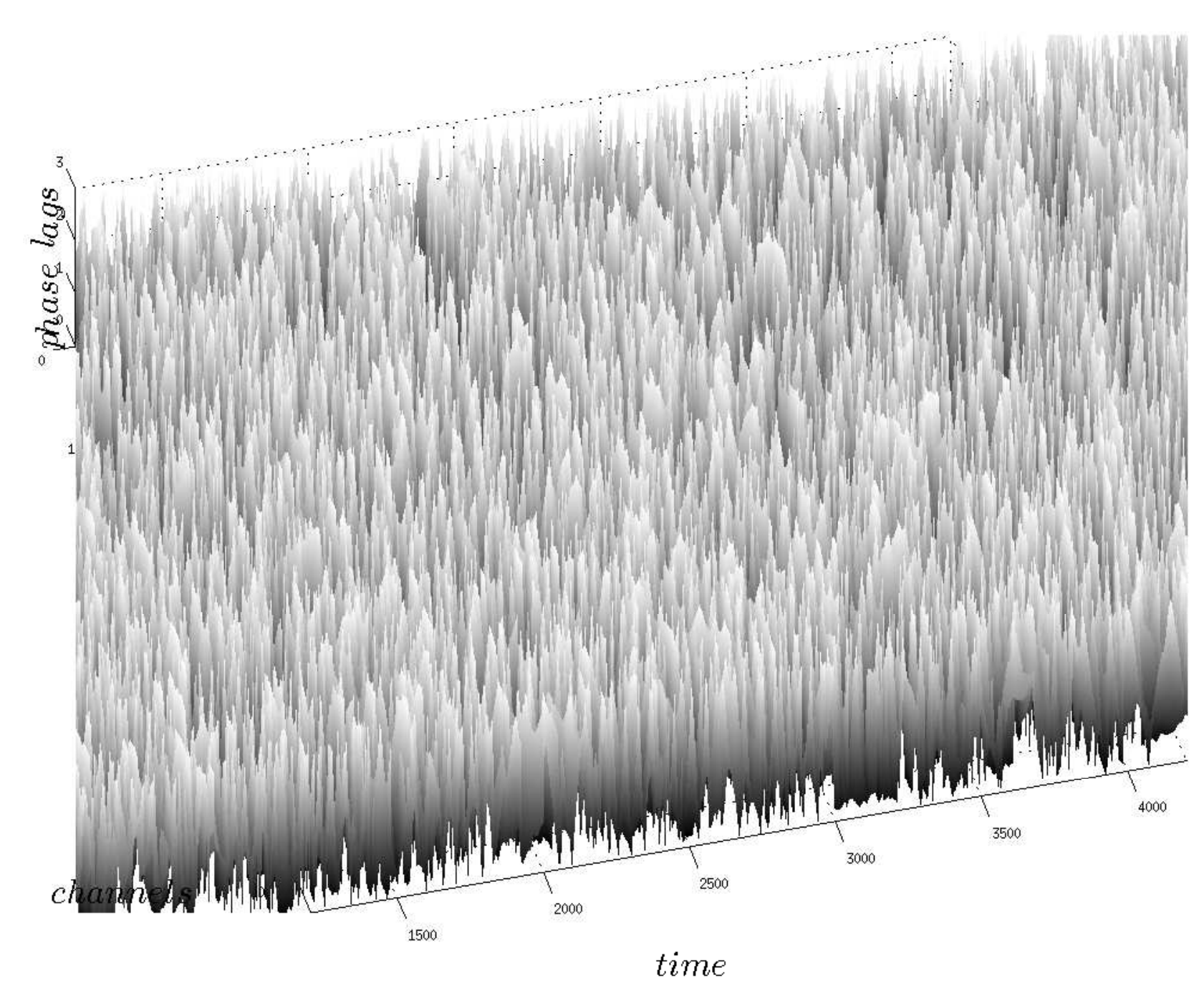} }
\subfigure[]{ \label{phaseLagsSS} \includegraphics[width=0.9\textwidth, height=0.5\textwidth]{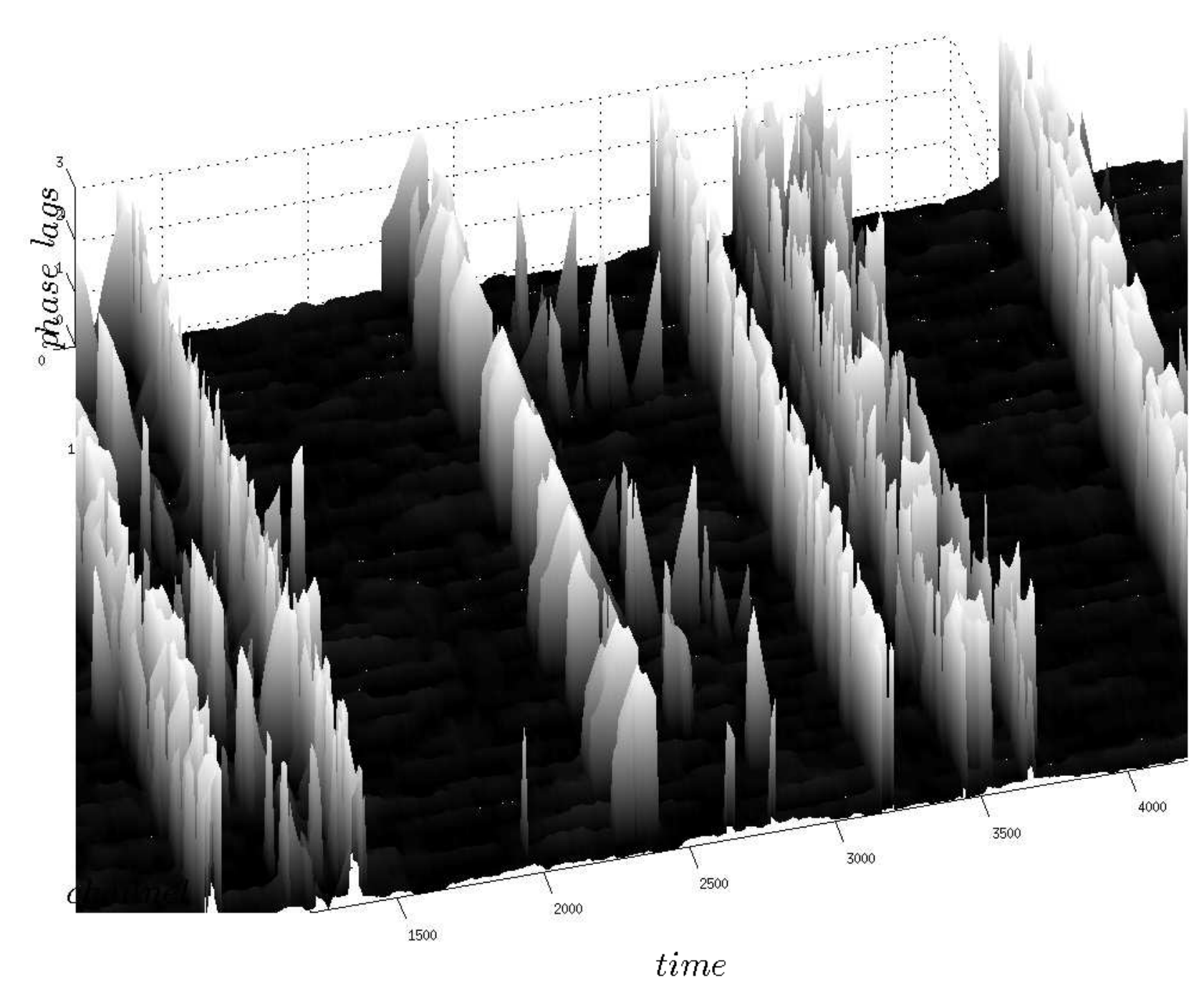} }
\caption{\label{phaseLagsNSphaseLagsSS}
Desynchronization to synchronization transitions; (a) at \(\omega\) below critical value there is no synchronization; (b) when \(\omega\) exceeds a critical value, the graph exhibits intermittent large-scale synchronization.
}
\end{center}
\end{figure}

\begin{figure}[ht!]
\begin{center}
\subfigure[]{ \label{nolearn1} \includegraphics[width=0.9\textwidth, height=0.5\textwidth]{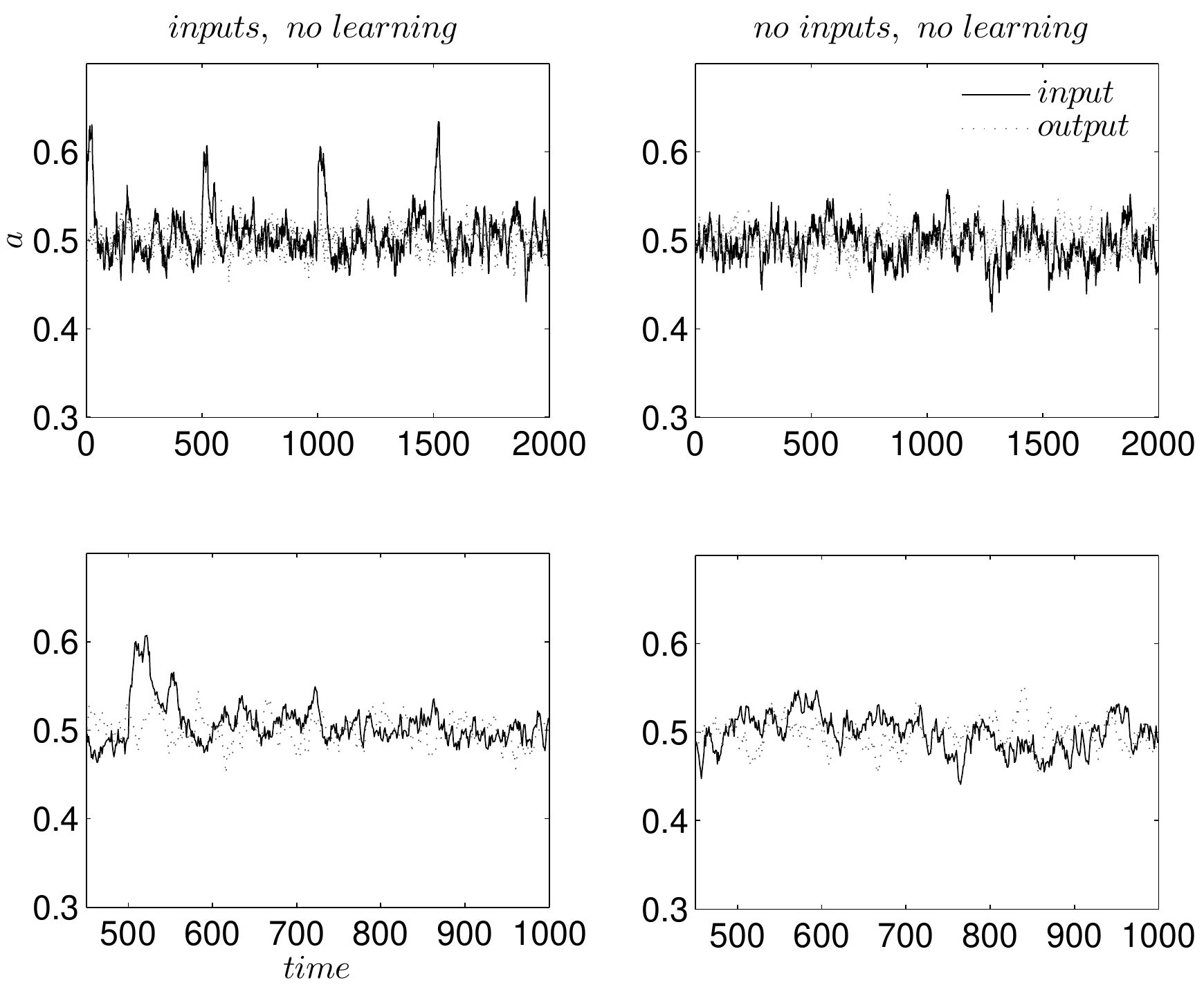} }
\subfigure[]{ \label{learn2} \includegraphics[width=0.9\textwidth, height=0.5\textwidth]{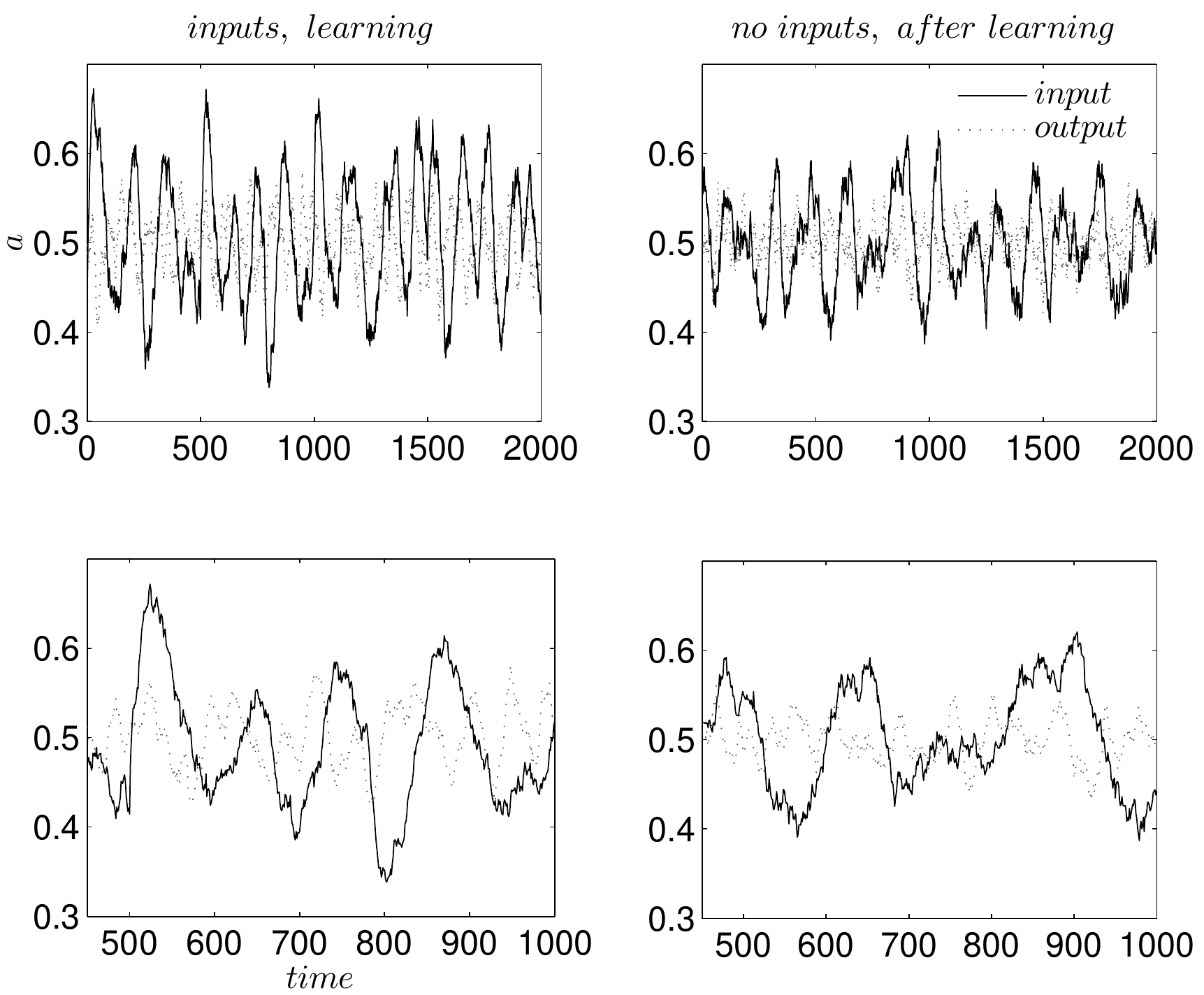} }
\caption{\label{learning}
Learning effect in coupled oscillators. (a) Activity levels without learning; first row contains extended time series plots, 2nd row has zoomed in sections. Input spike is shown at every 500 steps. The activity returns to base level after the input ceases.
\ref{learn2}: Activity levels with learning;The oscillations are prominent during learning and maintained even after the input step is removed (decayed). 
}
\end{center}
\end{figure}

\subsection{Intermittent Synchronization of Oscillations in 3 Coupled Double Arrays}

Finally, we study a system made of three coupled oscillators, each with its own inherent frequency. Such a system produces broad-band oscillations with intermittent synchronization-desynchronization behavior. In order to analyze quantitatively the synchronization features of this complex system with over 60,000 nodes in six layers, we merge the channels into a single two dimensional array at the readout. 
These channels will be used to measure the synchrony among the graph components.
First we find the dominant frequency of the entire graph using Fourier transform after ensemble averaging across the \(\approx\)10,000 channels for the duration of the numeric experiment. We set a time window \(W\) for the duration of two periods determined by the dominant frequency. We calculate the correlation between each channel and the ensemble average over this time window \(W\). At each time step, each channel is correlated with the ensemble over the window \(W\). Finally, we determine the dominant frequency of each channel and for the ensemble average at time \(t\) and the phase shifts are calculated. The difference in phases of dominant frequencies measure the phase lag between the channel and the ensemble. 

We plot the phase lags across time and space to visualize the synchrony between graph components, see Fig. \ref{phaseLagsNSphaseLagsSS}.  We can observe intermittent synchronization spatio-temporal patterns as the function of the \(\omega\), which represents the noise component of the evolution rule. For low \(\omega\) values the phase pattern is unstructured, see Fig. \ref{phaseLagsNS}. At a threshold \(\omega\) value intermittent synchronization occurs Fig. \ref{phaseLagsSS}. Further increasing \(\omega\), we reach  a regime with highly synchronized channels. 
Intermittent synchronization across large cortical areas has been observed in ECoG and EEG experiments as well. The correspondence between our theoretical/computational results and experimental data is encouraging. Further detailed studies are needed to identify the significance of the findings.

\subsection{Hebbian Learning Effects}

According to principles of neurodynamics, learning reinforces Hebbian cell assemblies which respond to known stimuli by destabilizing the broad-band chaotic dynamics an leading to a attractor basin through a narrow-band oscillation using gamma band career frequency. To test this model, we implemented learning in the neuropercolation model We use Hebbian learning, i.e., the weight from node i to node j is incrementally increased if these two nodes have the same state \((1-1\) or \(0-0)\). The weight from\( i\) to\(j \)j incrementally decreases if the two nodes have different activations \((0-1\) or \(1-0)\). Learning has been maintained during the 20 step periods when input was introduced. 

In the basal mode without learning, the 3 double layers can generate broad-band chaotic oscillations, see Fig\ref{nolearn1} where the lower row is zoomed-in version of top row. The spikes in Fig.\ref{nolearn1} indicate the presence of input signals. Inputs have been implemented by flipping a few \(\%\) of the input layer nodes to state Ô1Õ (active) for the duration of 20 iteration steps. During the learning stage, inputs are introduced 40 times (20 steps each), at regular intervals of 500 iteration steps. Without learning, the activity returns to the low-level chaotic state soon after the input ceases. 
Fig.\ref{learn2} shows that a narrow-band oscillation becomes prominent during learning, when a specific input is presented. After learning, the oscillatory behavior of the lattice dynamics is more prominent, even without input, but the learnt input elicits much more significant oscillations. This is the manifestation of the 6th and 7th principles of FreemanÕs neurodynamics, and it can be used to implement classification and control tasks using the neuropercolation model. 

Further steps, toward the 8th and 9th principles involve the analysis of the amplitude modulation (AM) pattern, which is converted to a feature vector of dimension \(n\). Here \(n\) is the number of nodes in our lattice which, after some course-graining, correspond to the EEG/ECoG electrodes. They provide our observation window into the operation of the cortex or cortex simulation. The feature vector is used to describe or label the cortical state at any moment. The AM pattern is formed by a low-dimensional, quasi-limit cycle attractor after synaptic changes with learning. The synaptic weights are described in a matrix, and the combination of different modalities to form Gestalts is done by concatenation of feature vectors. Research into this direction is in progress.

\section{Conclusions}

We employed neuropercolation model to demonstrate basic principles of neurodynamics.
Neuropercolation is a family of stochastic models based on the mathematical theory of probabilistic cellular automata on lattices and random graphs and motivated by structural and dynamical properties of neural populations. The existence of phase transitions has been demonstrated in probabilistic cellular automata and percolation models. Neuropercolation extends the concept of phase transitions to large interactive populations of nerve cells near criticality. Criticality in the neuropil is characterized by a critical region instead of a singular critical point. The trajectory of the brain as a dynamical systems crosses the critical region from less organized (gaseous) phase to more organized (liquid) phase during input induced destabilization and vise versa. Neuropercolation simulates these important results from mesoscopic ECOG/EEG recording across large spacial and temporal scales.

We showed that neuropercolation is able to naturally model input-induced destabilization of the cortex and consequent phase transitions due to sensory input and learning effects. Broad-band activity with scale-free power spectra is observed for unlearnt conditions. On the other hand, Hebbian learning results in the onset of narrow-band oscillations indicating the selection of specific learnt inputs. These observations demonstrate the operation of the first 7 building blocks of neurodynamics. Our results indicate that the introduced Hebbian learning effect can be used to identify and classify inputs. Considering that our model has 6 coupled lattices each with up to 10,000 nodes, massive computational resources have been involved in the simulations. Parallel chip implementation may be a natural way to expand the research in the future, especially when extending the work for multi-sensory Gestalt formation in the cortex.

\bibliographystyle{plain}
\bibliography{bibliography}
\end{document}

%% file: majorityRuleS.tex
\begin{picture}(200,95)(0,-5)
\put(33,50){\line(1,1){15}} \put(45,67){\small\(\omega_{x,x}\)}
\put(35,80){\circle*{10}} \put(36,87){\(\bf a\)}
\put(35,75){\vector(0,-1){20}} \put(18,68){\small\(\omega_{a,x}\)}
\put(5,50){\circle{10}} \put(4,57){\(\bf b\)}
\put(10,50){\vector(1,0){20}} \put(12,54){\small\(\omega_{b,x}\)}
\put(65,50){\circle{10}} \put(66,57){\(\bf d\)}
\put(60,50){\vector(-1,0){20}} \put(18,31){\small\(\omega_{c,x}\)}
\put(35,50){\circle*{10}} \put(39,42){\(\bf x\)}
\put(35,25){\vector(0,1){20}} \put(44,54){\small\(\omega_{d,x}\)}
\put(35,20){\circle*{10}} \put(38,26){\(\bf c\)}
\put(15,0){\(t-1\)}

\put(85,85){\(ax+:(a\to x)\to0\ w.p.\ \omega_{a,x}\)}
\put(85,72){\(bx-:(b\to x)\to0\ w.p.\ \omega_{b,x}\)}
\put(85,59){\(cx+:(c\to x)\to0\ w.p.\ \omega_{c,x}\)}
\put(85,46){\(dx+:(d\to x)\to1\ w.p.\ \omega_{d,x}\)}
\put(85,33){\(xx+:(x\to x)\to0\ w.p.\ \omega_{x,x}\)}
\put(85,10){\(x\ follows\ majority\ more \ likely\)}

\end{picture}

%% file: subgraphFig.tex
\begin{picture}(200,80)(0,0)

\multiput(15,15)(20,0){4}{\circle*{10}}
\multiput(15,35)(0,20){3}{\circle*{10}}
\multiput(35,35)(20,0){3}{\circle*{10}}
\multiput(35,55)(0,20){2}{\circle*{10}}
\multiput(55,55)(20,0){2}{\circle*{10}}
\multiput(55,75)(20,0){2}{\circle{10}}

\multiput(5,15)(0,20){3}{\line(1,0){80}} \put(5,75){\line(1,0){45}} \put(60,75){\line(1,0){10}} \put(80,75){\line(1,0){5}}
\multiput(15,5)(20,0){2}{\line(0,1){80}} \multiput(55,5)(20,0){2}{\line(0,1){65}} \multiput(55,80)(20,0){2}{\line(0,1){5}}
\qbezier(10,20)(11,19)(12,18) \qbezier(30,20)(31,19)(32,18) \qbezier(50,20)(51,19)(52,18) \qbezier(70,20)(71,19)(72,18)
\qbezier(10,40)(11,39)(12,38) \qbezier(30,40)(31,39)(32,38) \qbezier(50,40)(51,39)(52,38) \qbezier(70,40)(71,39)(72,38)
\qbezier(10,60)(11,59)(12,58) \qbezier(30,60)(31,59)(32,58) \qbezier(50,60)(51,59)(52,58) \qbezier(70,60)(71,59)(72,58)
\qbezier(10,80)(11,79)(12,78) \qbezier(30,80)(31,79)(32,78) \qbezier(50,80)(51,79)(52,78) \qbezier(70,80)(71,79)(72,78)

\multiput(115,15)(20,0){4}{\circle*{10}}
\multiput(115,35)(0,20){3}{\circle*{10}}
\multiput(135,35)(20,0){3}{\circle*{10}}
\multiput(135,55)(0,20){2}{\circle*{10}}
\multiput(155,55)(20,0){2}{\circle*{10}}
\multiput(155,75)(20,0){2}{\circle{10}}

\put(105,15){\line(1,0){50}} \put(170,15){\vector(-1,0){10}} \put(180,15){\line(1,0){5}} \multiput(105,35)(0,20){2}{\line(1,0){80}} \put(105,75){\line(1,0){45}} \put(170,75){\vector(-1,0){10}} \put(180,75){\line(1,0){5}} \put(157,16){\vector(1,4){14}} \put(159.5,71.5){\vector(1,-4){13}}
\multiput(115,5)(20,0){2}{\line(0,1){80}} \multiput(155,5)(20,0){2}{\line(0,1){65}} \multiput(155,80)(20,0){2}{\line(0,1){5}}
\qbezier(110,20)(111,19)(112,18) \qbezier(130,20)(131,19)(132,18) \qbezier(150,20)(151,19)(152,18) \qbezier(170,20)(171,19)(172,18)
\qbezier(110,40)(111,39)(112,38) \qbezier(130,40)(131,39)(132,38) \qbezier(150,40)(151,39)(152,38) \qbezier(170,40)(171,39)(172,38)
\qbezier(110,60)(111,59)(112,58) \qbezier(130,60)(131,59)(132,58) \qbezier(150,60)(151,59)(152,58) \qbezier(170,60)(171,59)(172,58)
\qbezier(110,80)(111,79)(112,78) \qbezier(130,80)(131,79)(132,78) \qbezier(150,80)(151,79)(152,78) \qbezier(170,80)(171,79)(172,78)

\end{picture}

%% file: nDapprox.tex
\begin{picture}(205,100)(0,0)
\put(2,70){0} \qbezier(7,75)(8,76)(9,77) \put(18,70){1} \qbezier(23,75)(24,76)(25,77) \put(33,70){2} \qbezier(38,75)(39,76)(40,77) \put(48,70){3} \qbezier(53,75)(54,76)(55,77)
\put(2,55){4} \qbezier(7,60)(8,61)(9,62) \put(18,55){5} \qbezier(23,60)(24,61)(25,62) \put(33,55){6} \qbezier(38,60)(39,61)(40,62) \put(48,55){7} \qbezier(53,60)(54,61)(55,62)
\put(2,40){8} \qbezier(7,45)(8,46)(9,47) \put(18,40){9} \qbezier(23,45)(24,46)(25,47) \put(31,40){10} \qbezier(41,45)(42,46)(43,47) \put(46,40){11} \qbezier(56,45)(57,46)(58,47)
\put(1,25){12} \qbezier(11,30)(12,31)(13,32) \put(16,25){13} \qbezier(26,30)(27,31)(28,32) \put(31,25){14} \qbezier(41,30)(42,31)(43,32) \put(46,25){15} \qbezier(56,30)(57,31)(58,32)
\multiput(0,28)(0,15){4}{\line(-1,2){5}} \multiput(56,28)(0,15){2}{\line(1,-2){5}} \multiput(54,58)(0,15){2}{\line(1,-2){5}}
\multiput(10,73)(15,0){3}{\line(1,0){5}} \multiput(10,58)(15,0){3}{\line(1,0){5}} \multiput(10,43)(15,0){3}{\line(1,0){5}} \multiput(10,28)(15,0){3}{\line(1,0){5}}
\multiput(4,18)(0,15){5}{\line(0,1){5}} \multiput(20,18)(0,15){5}{\line(0,1){5}} \multiput(35,18)(0,15){5}{\line(0,1){5}} \multiput(50,18)(0,15){5}{\line(0,1){5}}
\put(155,88){0} \qbezier(160,93)(161,94)(162,95) \qbezier(161,90)(165,88)(169,85) \qbezier(158,87)(152,67)(186,46)
\put(170,81){1} \qbezier(175,86)(176,87)(177,88) \qbezier(174,83)(179,79)(182,76) \qbezier(172,80)(160,55)(180,31)
\put(180,69){2} \qbezier(185,75)(186,76)(187,77) \qbezier(184,68)(185,66)(186,63) \qbezier(182,68)(162,45)(172,22)
\put(185,55){3} \qbezier(190,60)(191,61)(192,62) \qbezier(188,53)(188,51)(188,48) \qbezier(185,55)(160,40)(158,14)

\put(185,40){4} \qbezier(190,45)(191,46)(192,47) \qbezier(187,39)(186,36)(184,33) \qbezier(184,40)(154,40)(144,14)
\put(180,26){5} \qbezier(185,31)(186,32)(187,33) \qbezier(181,25)(177,21)(175,19) \qbezier(179,27)(154,40)(130,19)
\put(170,14){6} \qbezier(174,21)(175,22)(176,23) \qbezier(170,13)(165,12)(161,11) \qbezier(169,16)(155,35)(122,29)
\put(155,7){7} \qbezier(160,12)(161,13)(162,14) \qbezier(153,10)(150,10)(147,10) \qbezier(155,14)(142,38)(116,42)

\put(140,7){8} \qbezier(145,12)(146,13)(147,14) \qbezier(139,10)(135,12)(131,14) \qbezier(141,14)(138,42)(117,56)
\put(125,14){9} \qbezier(129,21)(130,22)(131,23) \qbezier(123,19)(122,21)(120,24) \qbezier(127,21)(146,45)(120,68)
\put(112,26){10} \qbezier(121,33)(122,34)(123,35) \qbezier(114,34)(113,36)(113,38) \qbezier(119,34)(140,50)(131,80)
\put(106,40){11} \qbezier(113,47)(114,49)(115,50) \qbezier(111,47)(111,50)(111,53) \qbezier(115,47)(150,60)(144,87)

\put(106,55){12} \qbezier(114,62)(115,63)(116,64) \qbezier(112,62)(113,65)(114,67) \qbezier(116,61)(139,62)(155,87)
\put(112,69){13} \qbezier(121,76)(122,77)(123,78) \qbezier(118,77)(121,80)(124,82) \qbezier(122,74)(147,63)(170,82)
\put(125,81){14} \qbezier(130,88)(131,89)(132,90) \qbezier(135,88)(137,89)(139,90) \qbezier(135,84)(148,65)(179,71)
\put(140,88){15} \qbezier(145,95)(146,96)(147,97) \qbezier(150,90)(152,90)(153,90) \qbezier(147,87)(152,54)(184,58)
\end{picture}

%% file: modelStructure.tex
\begin{picture}(205,40)
\put(0,19){\line(1,0){2}} \multiput(10,19)(12,0){9}{\line(1,0){5}} \multiput(118,19)(12,0){6}{\line(1,0){4}} \put(192,19){\line(1,0){2}}
\put(4,17){\small 0} \put(16,17){\small 1} \put(28,17){\small 2} \put(40,17){\small 3} \put(52,17){\small 4} \put(64,17){\small 5} \put(76,17){\small 6} \put(88,17){\small 7} \put(100,17){\small 8} \put(112,17){\small 9} \put(121,17){\small 10} \put(133,17){\small 11} \put(145,17){\small 12} \put(157,17){\small 13} \put(169,17){\small 14} \put(181,17){\small 15}
\multiput(6,23)(12,0){10}{\line(0,1){3}} \multiput(125,23)(12,0){6}{\line(0,1){3}}
\qbezier(0,25)(2,25)(3,23) \qbezier(0,30)(14,32)(27,23) \qbezier(8,24)(29,40)(50,24) \qbezier(32,24)(53,40)(74,24) \qbezier(56,24)(77,40)(98,24) \qbezier(80,24)(101,40)(122,24) \qbezier(104,24)(125,40)(146,24) \qbezier(128,24)(149,40)(170,24) \qbezier(152,24)(173,40)(191,25) \qbezier(176,24)(182,34)(191,31)
\qbezier(0,12)(4,11)(15,16) \qbezier(0,8)(10,5)(38,16) \qbezier(18,16)(41,0)(62,16) \qbezier(43,16)(65,0)(86,16) \qbezier(68,16)(89,0)(112,16) \qbezier(92,16)(113,0)(135,16) \qbezier(116,16)(137,0)(159,16) \qbezier(140,16)(161,0)(183,16) \qbezier(163,16)(179,5)(191,8) \qbezier(187,16)(189,11)(191,10)
\end{picture}

%% file: generalModel.tex
\begin{picture}(205,60)
\put(0,45){0-D} \put(45,45){1-D} \put(90,45){2-D} \put(135,45){3-D} \put(160,45){...} \put(180,45){n-D}
\multiput(10,32)(45,0){4}{\line(0,1){3}}
\multiput(45,30)(45,0){3}{\line(1,0){20}}
\multiput(92,23)(45,0){2}{\line(1,1){15}}
\put(137,37){\line(1,-1){15}}
\multiput(10,30)(45,0){4}{\circle*{6}} \put(160,30){...}
\put(5,5){0} \put(50,5){\(|V|^{0}\)} \put(95,5){\(|V|^{\frac{1}{2}}\)} \put(140,5){\(|V|^{\frac{2}{3}}\)} \put(180,5){\(|V|^{\frac{n-1}{n}}\)}
\end{picture}

%% file: oscillatorEG2.tex
\begin{picture}(220,65)
\put(140,55){\(v\in G_{0}\) excites \(G_{1}\)} \put(140,45){\(v\to G_{1}\ 1\ if\ 1\ else\ 0\)}
\put(140,10){\(v\in G_{1}\) inhibits \(G_{0}\)} \put(140,0){\(v\to G_{0}\ 1\ if\ 0\ else\ 1\)}

\put(0,50){..} \multiput(15,50)(12,0){6}{\circle*{6}} 
\put(89,50){..} \put(111,50){\circle{6}} \put(123,50){\circle*{6}} \put(133,50){..}
\multiput(15,52)(12,0){6}{\line(0,1){3}} 
\multiput(111,52)(12,0){2}{\line(0,1){3}}
\put(8,50){\line(1,0){5}} 
\put(30,50){\line(1,0){30}} 
\multiput(54,50)(12,0){3}{\line(1,0){6}} \multiput(102,50)(12,0){3}{\line(1,0){6}}
\multiput(10,45)(12,0){4}{\line(1,1){10}} \qbezier(64,52)(66,54)(68,56) \put(70,45){\line(1,1){10}} \qbezier(113,52)(115,54)(117,56) \put(118,45){\line(1,1){10}}

\put(0,5){..} 
\multiput(15,5)(12,0){6}{\circle*{6}} \put(89,5){..} \multiput(111,5)(12,0){2}{\circle*{6}} \put(133,5){..}
\multiput(14,7)(12,0){6}{\line(0,1){3}} 
\multiput(110,7)(12,0){2}{\line(0,1){3}}
\multiput(6,5)(12,0){5}{\line(1,0){6}} 
\put(77,5){\line(1,0){5}} \multiput(102,5)(12,0){3}{\line(1,0){6}}
\qbezier(10,0)(12,2)(14,4) 
\multiput(22,0)(12,0){5}{\line(1,1){10}} \qbezier(106,0)(108,2)(110,4) \put(118,0){\line(1,1){10}}

\put(16,48){\vector(0,-1){41}} \put(30,50){\line(1,-1){43}} \put(61,48){\line(0,-1){41}} \put(113,48){\vector(0,-1){41}}
\put(17,4){\vector(2,1){90}} \put(110,3){\vector(-2,1){92}}

\end{picture}

%% file: oscillator2EG2.tex
\begin{picture}(220,150)
\put(185,135){.. \(G_{0}\)} \put(185,95){.. \(G_{1}\)}
\put(0,135){..} \multiput(15,135)(12,0){6}{\circle*{6}} 
\put(89,135){..} \put(111,135){\circle{6}} 
\multiput(123,135)(12,0){5}{\circle*{6}}
\multiput(15,137)(12,0){6}{\line(0,1){3}} 
\multiput(111,137)(12,0){6}{\line(0,1){3}} 
\put(8,135){\line(1,0){5}} 
\multiput(30,135)(12,0){5}{\line(1,0){6}} \multiput(102,135)(12,0){7}{\line(1,0){6}} 
\multiput(10,130)(12,0){6}{\line(1,1){10}} 
\qbezier(113,137)(115,139)(117,141) 
\multiput(118,130)(12,0){5}{\line(1,1){10}}

\put(0,95){..} 
\multiput(15,95)(12,0){6}{\circle*{6}} \put(89,95){..} \multiput(111,95)(12,0){6}{\circle*{6}} 
\multiput(14,97)(12,0){5}{\line(0,1){3}} 
\multiput(110,97)(12,0){6}{\line(0,1){3}} 
\multiput(6,95)(12,0){7}{\line(1,0){6}} 
\multiput(102,95)(12,0){7}{\line(1,0){6}} 
\qbezier(10,90)(12,92)(14,94) 
\put(22,90){\line(1,1){10}} 
\multiput(22,90)(12,0){5}{\line(1,1){10}} 
\multiput(106,90)(12,0){5}{\line(1,1){10}} 
\qbezier(172,97)(174,99)(176,101)

\put(16,133){\line(0,-1){37}}
\put(75,95){\line(0,-1){48}}

\put(185,45){.. \(G_{2}\)} \put(185,5){.. \(G_{3}\)}
\put(0,45){..} \multiput(15,45)(12,0){6}{\circle*{6}} 
\put(89,45){..} 
\multiput(111,45)(12,0){6}{\circle*{6}}
\multiput(15,47)(12,0){5}{\line(0,1){3}} 
\multiput(111,47)(12,0){6}{\line(0,1){3}} 
\multiput(6,45)(12,0){7}{\line(1,0){6}} 
\multiput(102,45)(12,0){7}{\line(1,0){6}} 
\multiput(11,40)(12,0){4}{\line(1,1){10}} \qbezier(64,47)(66,49)(68,51) \put(70,40){\line(1,1){10}} \multiput(106,40)(12,0){4}{\line(1,1){10}} \qbezier(154,39)(156,41)(158,43) \put(166,40){\line(1,1){10}}

\put(0,5){..} 
\multiput(15,5)(12,0){6}{\circle*{6}} 
\put(89,5){..} 
\multiput(111,5)(12,0){6}{\circle*{6}} 
\multiput(15,7)(12,0){6}{\line(0,1){3}} 
\multiput(110,7)(12,0){6}{\line(0,1){3}} 
\multiput(6,5)(12,0){5}{\line(1,0){6}} \put(55,5){\line(1,0){5}} 
\put(77,5){\line(1,0){5}} 
\multiput(102,5)(12,0){7}{\line(1,0){6}} 
\qbezier(10,0)(12,2)(14,4) 
\multiput(22,0)(12,0){5}{\line(1,1){10}} \multiput(106,0)(12,0){6}{\line(1,1){10}} 

\put(157,45){\line(-1,2){44}}
\put(17,4){\line(1,1){44}} \put(60,5){\line(-1,4){32}} \qbezier(76,5)(140,24)(169,94)

\end{picture}